\documentclass{aa}
\usepackage{graphicx}
\usepackage{txfonts}

\newcommand{\msun}{\ensuremath{M_\odot}}                          
\newcommand{\rsun}{\ensuremath{R_\odot}}                          
\newcommand{\Msun}{\ensuremath{\,{\rm M}_\odot}}                  
\newcommand{\Rsun}{\ensuremath{\,{\rm R}_\odot}}                  
\newcommand{\lsun}{\ensuremath{L_\odot}}                          
\newcommand{\Teff}{\ensuremath{T_{\rm eff}}}                      
\newcommand{\Vsync}{\ensuremath{V_{\rm synch}}}                   
\newcommand{\EBV}{\ensuremath{E_{B-V}}}                           
\newcommand{\EVK}{\ensuremath{E_{V-K}}}                           
\newcommand{\Eby}{\ensuremath{E_{b-y}}}                           
\newcommand{\kms}{\,km\,s$^{-1}$}                                 
\newcommand{\mc}[1]{\multicolumn{2}{c}{#1}}                       
\newcommand{\Mbol}{\ensuremath{M_{\rm bol}}}                      
\newcommand{\sci}[2]{{:1}\!\times10^{:2}}                         
\newcommand{\Vsys}{\ensuremath{V_\gamma}}                         

\begin{document}

\title{Absolute dimensions of eclipsing binaries. XXIV.\thanks{Based on observations
carried out with the Str\"omgren Automatic Telescope (SAT) and the FEROS spectrograph
at ESO, La Silla, Chile (62.H-0319, 62.L-0284, 63.H-0080, 64.L-0031)}}

\subtitle{The Be star system DW\,Carinae, a member of the open cluster Collinder\,228}
\titlerunning{Absolute dimensions of the eclipsing binary DW\,Carinae}

\author{John Southworth\inst{1,2} \and J.\ V.\ Clausen\inst{1}}
\authorrunning{Southworth \and Clausen}

\offprints{J.\ Southworth}

\institute{Niels Bohr Institute, Copenhagen University, Juliane Maries Vej 30, DK-2100 Copenhagen \O, Denmark.
           \and Department of Physics, University of Warwick, Coventry, CV4 7AL, UK. \\
\email{jkt@astro.keele.ac.uk (JS), jvc@astro.ku.dk (JVC)}}

\date{Received ???? / Accepted ????}

\abstract
{The study of detached eclipsing binaries which are members of stellar clusters is a powerful way of determining the properties of the cluster and of constraining the physical ingredients of theoretical stellar evolutionary models.}
{DW\,Carinae is a close but detached early B-type eclipsing binary in the young open cluster Collinder\,228.
We have measured accurate physical properties of the components of DW\,Car (masses and radii to 1\%, effective temperatures to 0.02\,dex) and used these to derive the age, metallicity and distance of Collinder\,228.}
{The rotational velocities of both components of DW\,Car are high, so we have investigated the performance of double-Gaussian fitting, one- and two-dimensional cross-correlation and spectral disentangling for deriving spectroscopic radial velocites in the presence of strong line blending. Gaussian and cross-correlation analyses require substantial corrections for the effects of line blending, which are only partially successful for cross-correlation. Spectral disentangling is to be preferred because it does not assume anything about the shapes of spectral lines, and is not significantly affected by blending. However, it suffers from a proliferation of local minima in the least-squares fit. We show that the most reliable radial velocities are obtained using spectral disentangling constrained by the results of Gaussian fitting.
Complete Str\"omgren $uvby$ light curves have been obtained and accurate radii have been measured from them by modelling the light curves using the Wilson-Devinney program. This procedure also suffers from the presence of many local minima in parameter space, so we have constrained the solution using an accurate spectroscopic light ratio. The effective temperatures and reddening of the system have been found from Str\"omgren photometric calibrations.}
{The mass and radius of DW\,Car\,A are $M_{\rm A} = 11.34 \pm 0.12$\Msun\ and $R_{\rm A} = 4.558 \pm 0.045$\Rsun. The values for DW\,Car\,B are $M_{\rm B} = 10.63 \pm 0.14$\Msun\ and $R_{\rm B} = 4.297 \pm 0.055$\Rsun. Str\"omgren photometric calibrations give effective temperatures of ${\Teff}_{\rm A} = 27\,900 \pm 1000$\,K and ${\Teff}_{\rm B} = 26\,500 \pm 1000$\,K, and a reddening of $\Eby = 0.18 \pm 0.02$, where the quoted uncertainties include a contribution from the intrinsic uncertainty of the calibrations.
The membership of DW\,Car in Cr\,228 allows us to measure the distance, age and chemical composition of the cluster. We have used empirical bolometric corrections to calculate a distance modulus of $12.24 \pm 0.12$\,mag for DW\,Car, which is in agreement with, and more accurate than, literature values. A comparison between the properties of DW\,Car and the predictions of recent theoretical evolutionary models is undertaken in the mass--radius and mass--\Teff\ diagrams. The model predictions match the measured properties of DW\,Car for an age of about 6\,Myr and a fractional metal abundance of $Z \approx 0.01$.}
{}

\maketitle

\section{Introduction}                                                                       \label{sec:intro}

The study of detached eclipsing binary star systems (dEBs) is one of the few ways in which the absolute properties of stars can be measured directly and accurately (Andersen \cite{andersen91}). These objects are therefore of fundamental importance to the understanding of the characteristics of single stars. The masses and radii of the component stars in dEBs can be determined to accuracies of better than 1\% from the analysis of light curves and double-lined radial velocity curves (e.g., Southworth et al.\ \cite{SMSe}; Torres et al.\ \cite{torres97}). The effective temperatures of the two stars can be found by spectral analysis or photometric calibrations, allowing the luminosities and distance of the stars to be calculated directly (e.g., Southworth, Maxted \& Smalley \cite{SMSd}; Clausen \& Gim\'enez \cite{clausen91}).

One of the most important uses of the physical properties of dEBs is as a check of the predictions of theoretical models of stellar evolution. The two components of a dEB have the same age and chemical composition, as they were born together, but in general different masses, radii and luminosities. Theoretical models must therefore be able to match their accurately-known properties for one age and chemical composition. A good match is often achieved (e.g., Southworth, Maxted \& Smalley \cite{SMSb}) for two reasons. Firstly, the models are in general fairly reliable, and the effects of improved input physics lead to only minor adjustments in the evolutionary phases for which we are able to study dEBs. Secondly, there are several quantities which are not in general independently known for dEBs, including age, metal abundance and helium abundance, and these can be freely adjusted to help the models match the observed properties of a dEB. This last point means that it is extremely difficult to investigate the success of the treatment in theoretical models of several physical phenomena, for example convective core overshooting, mixing length, mass loss and opacity (see Cassisi \cite{cassisi04}, \cite{cassisi05}).

Extra constraints on the properties of a dEB can be obtained if it is a member of a co-evolutionary group of stars (Southworth, Maxted \& Smalley \cite{SMSa}; Hebb, Wyse \& Gilmore \cite{hebb04}). If the age and chemical composition of the stars is known independently then these quantities can be used to constrain the models, making it more challenging for them to match observations and so allowing the effects of more subtle physical phenomena to be investigated (e.g., Torres \& Ribas \cite{torres02}). If the age and composition of the stellar group is not known then these may be found very precisely by comparing model predictions to the properties of the dEB (Southworth et al.\ \cite{SMSa}, \cite{SMSb}; Thompson et al.\ \cite{oglegc17}). In addition, dEBs are excellent distance indicators (e.g., Guinan et al.\ \cite{guinan98}; Clausen \cite{clausen04}), allowing accurate and precise distances to be found to the parent stellar systems in several empirical and semi-empirical ways (Southworth et al.\ \cite{SMSd}; Guinan et al.\ \cite{guinan98}).

\subsection{DW Carinae}

Here we present a study of the young, early B-type system \object{DW\,Car}, which is a member of the Southern open cluster \object{Collinder\,228} (Ferrer et al.\ \cite{ferrer85}). We have obtained complete and accurate Str\"omgren $uvby$ light curves, which are presented in Clausen et al.\ (\cite{clausen06}, hereafter Paper\,I) along with a revised orbital ephemeris. We have also obtained extensive spectroscopy using the FEROS \'echelle spectrograph and the 1.5\,m telescope at ESO La Silla.

DW\,Car (Table~\ref{table:dwcar}) is composed of two similar B1\,V stars in a 1.3 day orbit. The stars are well separated, despite the shortness of the orbital period, as they are essentially unevolved. Such a system is particularly useful because there are very few well-studied components of dEBs with masses of 10\Msun\ or greater which are close to the zero-age main sequence (ZAMS) (Andersen \cite{andersen91}; Gies \cite{gies03}). High-mass unevolved dEBs can be particularly useful for investigating the effects of the opacities used in theoretical models (Torres et al.\ \cite{torres97}).

\begin{table} \caption{Identifications and astrophysical data for DW\,Car. $T_0$ is a reference
time of mid-eclipse of the primary star.
\newline {\bf References:} (1) Cannon (\cite{cannon25}); (2) H{\o}g et al.\ (\cite{hog00}); (3)
Levato \& Malaroda (\cite{levato81}); (4) Paper\,I; quantities in parentheses are uncertainties
in the final digit of the quanities.} 
\label{table:dwcar} \centering
\begin{tabular}{lr@{}lc} \hline \hline
                              &     & DW\,Car               & References\\\hline
Henry Draper Catalogue        &     & HDE 305543            & 1         \\
Tycho number                  &     & TYC 8957-1314-1       & 2         \\
$\alpha_{2000}$               &     & 10 43 10.1            & 2         \\
$\delta_{2000}$               & $-$ & 60 02 12              & 2         \\[2pt]
Spectral type                 &     & B1\,V + B1\,V         & 3         \\[2pt]
$V$                           &     & 9.675 $\pm$ 0.005     & 4         \\
$b-y$                         &     & 0.067 $\pm$ 0.005     & 4         \\
$m_1$                         &     & 0.046 $\pm$ 0.008     & 4         \\
$c_1$                         & $-$ & 0.017 $\pm$ 0.008     & 4         \\
$\beta$                       &     & 2.528 $\pm$ 0.007     & 4         \\
$T_0$ (HJD)                   &     & 2\,446\,828.6692(4)   & 4         \\
Orbital period (days)         &     & 1.32774934(44)        & 4         \\
\hline \end{tabular} \end{table}

The eclipsing nature of DW\,Car was discovered by Hertzsprung (\cite{hertz24}), who measured the orbital period to be 0.66382\,d from a photographic light curve containing 251 observations. van den Hoven van Genderen (\cite{vdhvg39}) refined the orbital period using a 914-point photographic light curve, but stated that the period should be doubled even though there was no clear evidence that adjacent minima had different depths. Gaposchkin (\cite{gap52}, \cite{gap53}) published a photographic light curve containing over one thousand points and found an orbital period of 1.3277504\,d. His estimated masses and radii are quite close to the values we measure below.

A spectroscopic orbit for DW\,Car was published by Ferrer et al.\ (\cite{ferrer85}) based on 67 photographic observations. They found no evidence for orbital eccentricity, and a systemic velocity which is consistent with membership of Cr\,228. A spectral classification of B1\,V + B1\,V was provided by Levato \& Malaroda (\cite{levato81}).

We shall refer to the primary and secondary components of DW\,Car as star A and star B, respectively, where star A is eclipsed by star B at phase 0.0. Star A has a greater surface brightness and mass than star B.

\subsection{Collinder\,228}                                                                  \label{sec:cr228}

The open cluster Cr\,228 is a young and sparse cluster located in the Carina spiral arm feature. Its immediate surroundings are quite nebulous, making photometric studies very difficult.

Feinstein et al.\ (\cite{fein76}) found a reddening of $\EBV = 0.33 \pm 0.08$\,mag, an age of less than 5\,Myr and a distance modulus of $V_0 - M_V = 12.0 \pm 0.2$\,mag on the basis of a photoelectric $UBV$ study.

Th\'e, Bakker \& Antalova (\cite{the80}) obtained photoelectric observations in the Walraven system and found a distance modulus of 12.0\,mag assuming a normal reddening law (or 11.7\,mag assuming $R_V = \frac{A_V}{\EBV} =
4.0$).

Turner \& Moffat (\cite{turner80}) obtained photoelectric $UBV$ observations and showed that the reddening in the region was normal and that the open clusters Trumpler 14, Trumpler 15 and Cr\,228 are at a common distance modulus of $12.15 \pm 0.14$\,mag. They found that the age of Cr\,228 must be at least 1\,Myr from the presence of main sequence stars at $(B-V)_0 = -0.1$.

Tapia et al.\ (\cite{tapia88}) obtained $JHK$ photometry of the region, and found the reddening characteristics of Cr\,228, and other nearby clusters, to be anomalous. These authors found that the absolute visual extinction, $A_V$, is quite different when calculated from the colour excess \EBV\ rather than from colour excesses for longer-wavelength passbands. Under the assumption that \EBV\ is anomalous in this region, they found $A_V = 1.94 \pm 0.16$\,mag from \EVK, giving a distance modulus of $V_0 - M_V = 11.6 \pm 0.4$.

Carraro \& Patat (\cite{carraro01}) obtained $UBVRI$ photometry of Cr\,228 and several nearby clusters. They found an extinction of $\EBV = 0.30 \pm 0.05$\,mag, in agreement with Feinstein et al.\ (\cite{fein76}) but disagreeing with Tapia et al.\ (\cite{tapia88}). Their distance modulus, $11.4 \pm 0.2$\,mag, is smaller than that derived in most previous studies. Carraro \& Patat suggested that about 30\% of the stars in Cr\,228 are binary, in good agreement with the spectroscopic study of Levato et al.\ (\cite{levato90}).

Massey, DeGioia-Eastwood \& Waterhouse (\cite{massey01}) obtained new spectral types for 16 stars, and collected literature values for another 40 stars, in the region of Cr\,228. They found a median \EBV\ of 0.37 and a distance modulus of 12.5\,mag. They obtained the ages of the higher-mass unevolved stars from interpolation in effective temperature and absolute bolometric magnitude using the theoretical evolutionary tracks of Schaller et al.\ (\cite{schaller92}). The age and standard deviation is $2.2 \pm 1.0$\,Myr, but these authors note that the appearance of the HR diagram suggests a spread in ages in Cr\,228.

It is clear from the above discussion that the physical properties of the open cluster Cr\,228 are rather uncertain, because the cluster is quite sparse and located in a nebulous area. The brightest member of this cluster, \object{QZ\,Car} (HD\,9326), is a quadruple system which contains two spectroscopic binaries, one of which shows eclipses of a semi-detached nature (Mayer et al.\ \cite{mayer01}).


\section{Observations and data reduction}                                                      \label{sec:obs}

Light curves of DW\,Car in the Str\"omgren $uvby$ photometric system were obtained from the Str\"omgren Automated Telescope at ESO La Silla. These are presented in Paper\,I and contain 518 observations in each passband. Paper\,I also presents an updated orbital ephemeris and standard Str\"omgren $uvby$ and Crawford H$\beta$ indices for DW\,Car.

\subsection{Spectroscopic data}

29 high-resolution spectra were obtained using the FEROS fibre-fed \'echelle spectrograph on the ESO 1.52\,m telescope at La Silla (Kaufer et al.\ \cite{feros99}, \cite{feros00}) between 1998 November and 2000 January. This spectrograph is located in a temperature-controlled room and covers the spectral region from the Balmer jump to 8700\,\AA\ without interruption and at a constant velocity resolution of 2.7\kms\,px$^{-1}$ ($\lambda/\Delta\lambda=48000$). An observing log is given in Table~\ref{tab:dwcar_feros}.

A modified version of the MIDAS FEROS package, prepared by H.\ Hensberge, was used for the basic reduction\footnote{see these websites for further details: \\ {\scriptsize\tt http://www.ls.eso.org/lasilla/sciops/2p2/E2p2M/FEROS/DRS} \\ {\scriptsize\tt http://www.ls.eso.org/lasilla/sciops/2p2/E1p5M/FEROS/Reports}}. This package pays careful attention to background removal, definition and extraction of the individual orders and wavelength calibration, and provides a significant improvement on the standard quick-look reduction pipeline. The observations were reduced night by night using calibration exposures (thorium-argon and flat field) obtained during the preceding afternoon and following morning for each night. Standard (rather than optimal) extraction was applied, and no order merging was attempted. A standard error of the wavelength calibration of about 0.002--0.003\,\AA\ was typically obtained.

\begin{table} \caption{\label{tab:dwcar_feros} Logbook of the spectroscopic observations of
DW\,Car. HJD refers to the exposure midpoint, ID to the identification number, SD to if it
was included in the spectral disentangling analysis and $t_{\rm exp}$ to the exposure time.
The column headed `Obs' indicates the observer (H indicates Heidelberg/Copenhagen guaranteed
time observations). The signal to noise ratio, S/N, of single exposures was measured at
approximately 4000 and 5000\,\AA.}
\begin{tabular}{lccrrrr} \hline \hline
HJD minus         & ID      & SD & Obs    & Orbital & $t_{\rm exp}$ & S/N      \\
2\,400\,000       &         &    &        & phase   & (s)           &          \\ \hline
51144.83060       & s\,1    & y  & H      & 0.73512 &      600      &  35--60  \\ 
51145.84400       & s\,2    &    & H      & 0.49836 &      600      &  30--55  \\ 
51146.86755       & s\,3    & y  & H      & 0.26925 &      600      &  40--75  \\ 
51147.86649       & s\,4    &    & H      & 0.02161 &      600      &  35--60  \\ 
51148.80902       & s\,5    & y  & H      & 0.73148 &      600      &  40--85  \\ 
51149.87170       & s\,6    &    & H      & 0.53184 &      420      &  25--40  \\ 
51150.83746       & s\,7    & y  & H      & 0.25920 &      600      &  40--60  \\ 
51152.73657       & s\,8    &    & H      & 0.68953 &      600      &  25--60  \\ 
51172.69544       & s\,9    & y  & H      & 0.72164 &      900      &  50--80  \\ 
51174.78088       & s\,10   & y  & H      & 0.29229 &      900      &  75--110 \\ 
51176.80734       & s\,11   & y  & H      & 0.81853 &      600      &  65--95  \\ 
51178.84953       & s\,12   & y  & H      & 0.35661 &      600      &  70--90  \\ 
51179.87450       & s\,13   & y  & H      & 0.12857 &      600      &  65--100 \\ 
51180.81349       & s\,14   & y  & H      & 0.83578 &      600      &  60--80  \\ 
51181.87490       & s\,15   & y  & H      & 0.63519 &      600      &  65--90  \\ 
51182.86220       & s\,16   & y  & H      & 0.37877 &      600      &  60--100 \\ 
51184.86864       & s\,17   &    & H      & 0.88993 &      600      &  55--75  \\ 
51187.87364       & s\,18   &    & H      & 0.15316 &      600      &  35--65  \\ 
51189.86682       & s\,19   & y  & H      & 0.65433 &      600      &  50--75  \\ 
51190.84867       & s\,20   &    & H      & 0.39382 &      600      &  50--70  \\ 
51191.86279       & s\,21   & y  & H      & 0.15760 &      600      &  50--75  \\ 
51192.80846       & s\,22   & y  & H      & 0.86984 &      600      &  65--85  \\ 
51193.87953       & s\,23   &    & H      & 0.67652 &      720      &  65--95  \\ 
51197.85496       & s\,24   & y  & H      & 0.67063 &      600      &  40--65  \\ 
51199.87004       & s\,25   & y  & H      & 0.18830 &      900      &  60--85  \\ 
51209.72395       & s\,26   &    & JVC    & 0.60981 &     1800      & 110--140 \\ 
51382.51044       & s\,27   & y  & H      & 0.74468 &     1800      &  50--125 \\ 
51386.50761       & s\,28   &    & H      & 0.75517 &     1800      &  70--135 \\ 
51563.86655       & s\,29   & y  & JVC    & 0.33380 &     2400      & 105--150 \\ 
\hline \end{tabular} \end{table}


\section{Radial velocity analysis}                                                              \label{sec:rv}

The FEROS spectra of DW\,Car cover essentially the entire optical wavelength range, but are disappointingly short of stellar features. The only absorption lines which are easy to discern are the hydrogen Balmer lines and several He\,I lines. The high rotational velocity of both stars, approximately 170\kms, twinned with the fact that the spectrum of each star is diluted by the flux of the companion, means that only the strongest
spectral lines are visible.

\begin{figure} \resizebox{\hsize}{!}{\includegraphics{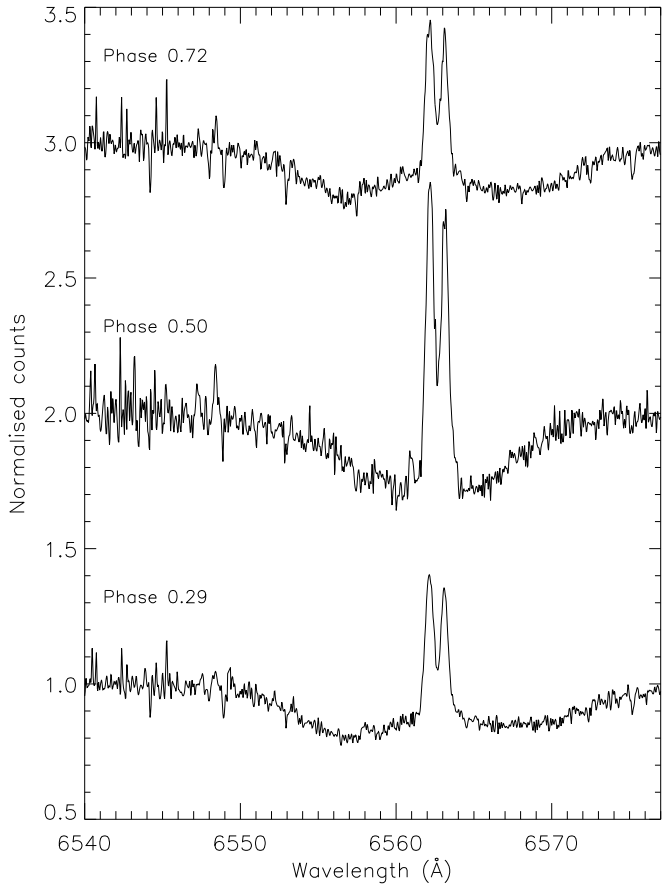}}
\caption{Continuum-normalised H$\alpha$ profiles of DW\,Car at phases
0.29, 0.50 and 0.72 (spectra s\,10, s\,2 and s\,9). Spectra s\,2 and
s\,9 have been shifted by +1 and +2, respectively, for clarity.}
\label{fig:halpha} \end{figure}

The H$\alpha$ line shows significant emission with an absorption superimposed on the line core. The emission and emission absorption do not follow the orbital motion of the stars but remain at the systemic velocity of the system. Slight emission is also visible in the centre of H$\beta$ for some of the spectra, depending on signal to noise ratio. Fig.~\ref{fig:halpha} shows H$\alpha$ spectra at phases 0.27, 0.50 and 0.72. Whilst the emission appears to be twice as strong at phase 0.50, this is simply an artefact of the continuum normalisation as the system is about half as bright at eclipse midpoint compared to during quadrature phases. Further investigation of the emission line strengths will require flux-calibrated spectra (D.\ Lennon,
private communication).

\begin{figure} \resizebox{\hsize}{!}{\includegraphics{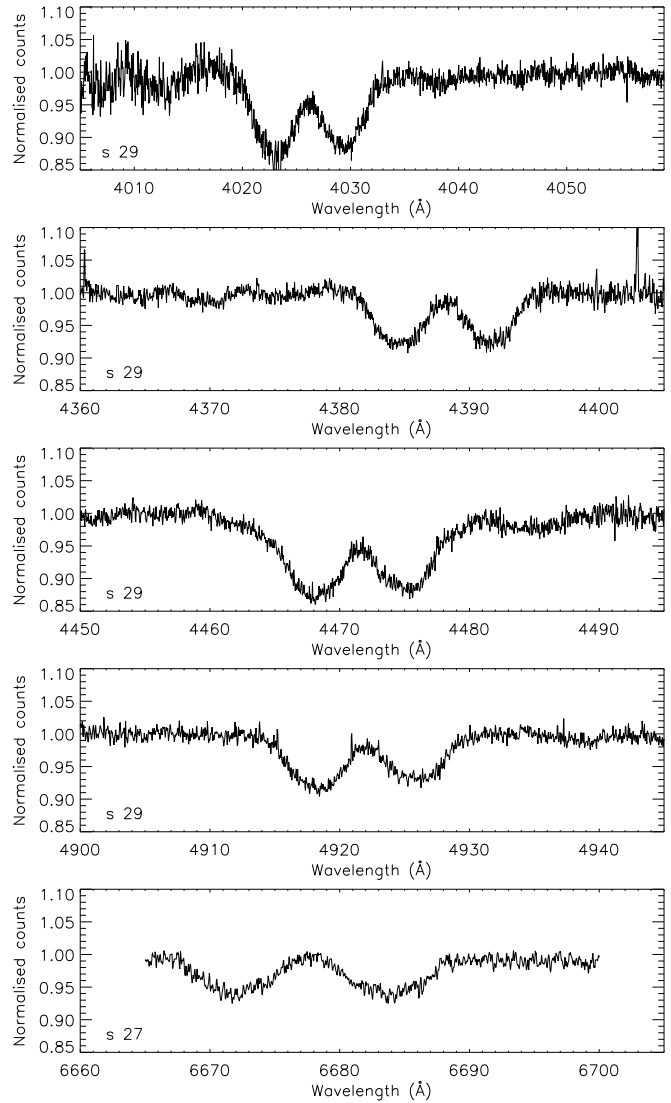}}
\caption{Continuum-normalised example spectra in the regions of the
five He\,I lines used in the radial velocity analysis. Spectrum s\,29
(phase 0.33) is shown in the top four panels and spectrum s\,27 (phase
0.74) in the last panel. Shallow lines can also be seen at 4009\,\AA\
(He\,I) and 4481 (Mg\,II). We have chosen spectra with a high signal to
noise ratio to plot -- most of the spectra have a significantly lower
ratio.} \label{fig:specplot} \end{figure}

Narrow regions around the five He\,I absorption lines at 4026, 4388, 4471, 4922 and 6678\,\AA\ were chosen for our radial velocity analysis. The remaining visible He\,I lines are all broad and very shallow, and we have found that minor errors in continuum placement can cause a large systematic error in the measured centres and widths of these lines. The hydrogen Balmer lines were not included in the radial velocity analysis, as line blending can cause large and systematic underestimation of the velocity semiamplitudes (Petrie, Andrews \& Scarfe \cite{petrie67}; Hilditch \cite{hilditch73}; Andersen \cite{andersen75}). Regions from an example spectrum around the lines used are shown in Fig.~\ref{fig:specplot}.

As there are very few spectral lines strong enough to yield reliable radial velocities, we have used four different methods to measure these velocities, allowing us to investigate which are the most reliable. The techniques are fitting double Gaussian functions (hereafter called {\sc 2gauss}), cross-correlation ({\sc onecor}), two-dimensional cross-correlation ({\sc todcor}) and spectral disentangling. Spectroscopic orbits were calculated for each spectral line and for each technique.

The {\sc onecor} and {\sc todcor} algorithms require a template spectrum which closely matches the spectral characteristics of the components of DW\,Car. We have used a FEROS spectrum of the rotational velocity standard star \object{HR\,1855} ($\upsilon$\,Ori, spectral type B0\,V), which has a rotational velocity of 10--20\kms\ (Lyubimkov, Rostopchin \& Lambert \cite{lyubimkov04}; Abt, Levato \& Grosso \cite{abt02}). This spectrum has signal to noise ratios of between 235 and 350 depending on the \'echelle order, and was broadened by a rotational velocity of 150\kms\ using the method of Gray (\cite{gray92}) with linear limb darkening coefficients of between 0.39 and 0.30 depending on the wavelength of the spectral order. We have measured the radial velocity of this star to be $+19.9 \pm 2.4$\kms\ from Gaussian fits to several He\,I lines, in agreement with the Wilson-Evans-Batten radial velocity catalogue (Duflot, Figon \& Meyssonnier \cite{duflot95}) value of $+17.4$\kms. We have subtracted 19.9\kms\ from all the velocities of DW\,Car measured using HR\,1855 as a template.

\subsection{Gaussian fitting ({\sc 2gauss})}                                             \label{sec:rv:2gauss}

The spectra of DW\,Car were velocity-binned to 5\kms\,px$^{-1}$ and a double Gaussian function was fitted to four of the five good He\,I lines (listed above) by the method of least squares with rejection of points deviating by more than 4\,$\sigma$ from an initial best fit. The fit to each spectrum was inspected by eye and poor fits were rejected. Double weights were assigned to a few fits which seemed to be of a particularly
high quality.

Spectroscopic orbits were fitted to the measured radial velocities for each spectral line and the systemic velocities of the two stars were not forced to be equal (see Popper \& Hill \cite{popper91}). The orbits were calculated using the method of Lehman-Filh\'es implemented in the {\sc sbop}\footnote{Spectroscopic Binary Orbit Program, \\ {\tt http://mintaka.sdsu.edu/faculty/etzel/}} program (Etzel \cite{etzel04}), which is a modified and expanded version of an earlier code by Wolfe, Horak \& Storer (\cite{wolfe67}). The ephemeris from Paper\,I was adopted (Table~\ref{table:dwcar}) and the orbit was assumed to be circular based on initial analyses, the shortness of the orbital period and the shape of the light curve.

The means and standard deviations of the full widths at half maximum of the fitted Gaussian functions are given in Table~\ref{table:specdata}.

\begin{table} \caption{Additional parameters found during the spectroscopic analysis. The
full widths at half maximum of the Gaussian functions come from the {\sc 2gauss} analysis
and light ratios ($\frac{\ell_{\rm B}}{\ell_{\rm A}}$) come from the {\sc todcor} analysis.}
\label{table:specdata} \centering
\begin{tabular}{l r@{$\pm$}l r@{$\pm$}l r@{$\pm$}l} \hline \hline
He\,I line  & \multicolumn{4}{c}{FWsHM of the Gaussian fits (\kms)} & \mc{Light ratio} \\
(\AA)       &       \mc{Star A}      &      \mc{Star B}             & \mc{from {\sc todcor}} \\ \hline
4026.189    & 171.5 &  9.6 & 161.9 & 11.2 & 0.839 & 0.138 \\
4387.928    & 167.8 & 10.1 & 163.8 &  6.7 & 0.855 & 0.106 \\
4471.681    & 194.9 & 16.2 & 200.8 & 21.6 & 0.893 & 0.059 \\
4921.929    & 173.9 &  9.2 & 163.8 & 10.6 & 0.856 & 0.082 \\
\hline \end{tabular} \end{table}

\subsection{One-dimensional cross-correlation ({\sc onecor})}

The velocity-binned spectra of DW\,Car were subjected to a cross-correlation analysis (Simkin \cite{simkin74}; Tonry \& Davis \cite{tonry79}) using Fast Fourier Transform techniques. We investigated using three different types of template spectra to derive radial velocities from the target spectra:%

\begin{itemize}%

\item We attempted to use spectrum s\,2 as a template, which was observed at conjunction (phase 0.498), because this provides the best match to the target spectra. These attempts failed because bad pixel values in the target spectra, caused by CCD defects, tended to correlate against their couterparts in the template spectrum and deform the cross-correlation function (CCF).%

\item The artificially broadened spectrum of HR\,1855 was used and found to give good results. This spectrum does not contain the bad pixels common to the spectra of DW\,Car. We therefore used this template to measure radial velocities.%

\item Synthetic spectra generated using the {\sc uclsyn} code (see Southworth et al.\ \cite{SMSa} for references). The results were inaccurate because the synthetic profiles were a poor match to the observations, as expected because {\sc uclsyn} is not intended for stars this hot. We did not pursue the use of synthetic spectra further because good results had already been obtained using template spectra of HR\,1855.%

\end{itemize}%

Radial velocities were accepted or rejected by eye, based on the shape of the CCF. The peaks of the CCF were then determined by least-squares quadratic interpolation and spectroscopic orbits were calculated using {\sc sbop} as before (Sect.~\ref{sec:rv:2gauss}). We experimented with using unbroadened template spectra as these should be less strongly affected by line blending, but found that the results were excessively noisy. This effect has been noted before and is due to the template and target spectra being quite different to each other.

\subsection{Two-dimensional cross-correlation ({\sc todcor})}                       \label{sec:rv:todcor}

The {\sc todcor} algorithm (Zucker \& Mazeh \cite{zucker94}) produces a two-dimensional CCF by cross-correlating two template spectra (one for each star) with each target spectrum, which should reduce the problems caused by blending effects when the stellar spectral lines are separated by less than their total broadening. As with the {\sc onecor} analysis, three types of template spectrum were investigated and the broadened spectrum of HR\,1855 was found to be the best. Radial velocities were measured by determining the position of the relevant peak of the CCF by fitting a surface with the IDL\footnote{\tt http://www.rsinc.com/idl/} procedure {\sc min-curve-surf}. They were accepted or rejected via visual inspection of the CCF, and orbits were calculated using {\sc sbop} as before.

The {\sc todcor} algorithm can be used to analytically evaluate the light ratio between the two component stars ($\frac{\ell_{\rm B}}{\ell_{\rm A}}$) on the basis of the relative strengths of their absorption lines (Zucker \& Mazeh \cite{zucker94}). The mean and standard deviation of the light ratio calculated for each spectrum is given in Table~\ref{table:specdata} for each \'echelle order. These light ratios are meaningful because of the negligible difference in spectral type between the two components of DW\,Car.

\subsection{Spectral disentangling}

\begin{figure} \resizebox{\hsize}{!}{\includegraphics{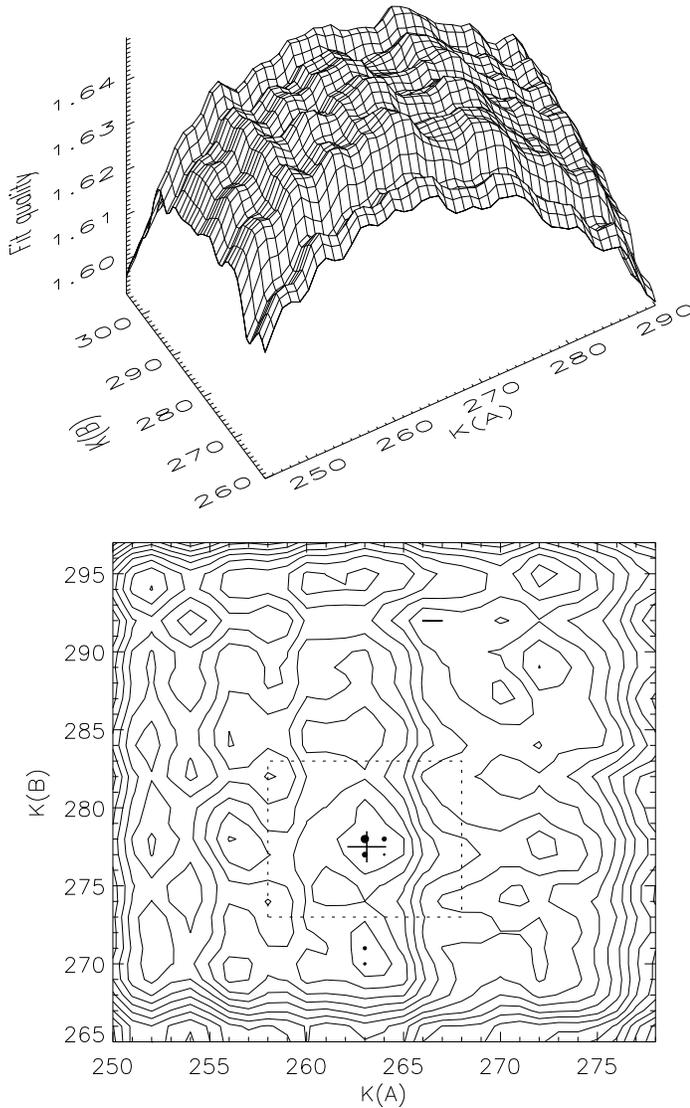}}
\caption{Variation of the quality of the fit for spectral disentangling
the \'echelle order 55 spectra of DW\,Car. The upper panel is a surface
plot of fit quality versus $K_{\rm A}$ and $K_{\rm B}$. The lower panel
is a contour plot of the same data with he six highest values of fit
quality indicated by filled circles (largest for the best fit and smallest
for the sixth-best fit). On the lower panel the area which a surface was
fitted to is indicated by a dashed box and the highest point of that
surface by a cross.} \label{fig:k1k2grid} \end{figure}

The spectral disentangling technique solves simultaneously for the contributions of the components to the observed composite spectra, and for the Doppler shifts in the component spectra. We have applied the method introduced by Simon \& Sturm (\cite{simon94}) using a version of the original code revised for the Linux operating system. It assumes a constant light level, so observations taken during eclipse were rejected from the analysis; DW\,Car is constant to within about 5\% outside eclipses. Dedicated IDL programs were applied to remove cosmic ray events and other defects, and for careful normalization of the individual orders. 

We have assumed the ephemeris from Paper\,I and a circular orbit, and fitted for the velocity semamplitudes of the spectroscopic orbit for each star, $K_{\rm A}$ and $K_{\rm B}$. Suitable spectral lines are present only in orders 55, 51, 50, 45 and 33, and these orders were analysed individually. The formal errors in $K_{\rm A}$ and $K_{\rm B}$ returned by the code are generally around 0.3\kms, which is much smaller than expected given the range in $K_{\rm A}$ and $K_{\rm B}$ for different orders (Table~\ref{table:rvorbits}, Table~\ref{table:rvorbits:cor}). It is not clear how to calculate robust uncertainties for spectroscopic orbits calculated in disentangling analyses (Hynes \& Maxted \cite{hynes98}), so we will not quote uncertainties for individual spectroscopic orbits.

We have also disentangled the spectra for a grid of fixed values of $K_{\rm A}$ (from 245 to 290\kms) and $K_{\rm B}$ (260 to 305\kms) with a spacing of 1\kms, to study how the quality of the least-squares fit varies. The results are shown in Fig.~\ref{fig:k1k2grid} for the spectra from \'echelle order 55. There are several local maxima but one clear global maximum. The highest point of a surface fitted by least squares around the global maximum is indicated. This diagram shows that whilst spectral disentangling can be used to determine spectroscopic orbits for binary stars, the best-fitting solution returned by minimisation algorithms cannot in general be trusted unless confirmed to be the global maximum by a grid search or by external information.

\subsection{Corrections from synthetic spectra}

The need to apply corrections to the radial velocities derived using the {\sc todcor} algorithm has been demonstrated in several cases, e.g., Torres et al.\ (\cite{torres97}) and Torres \& Ribas (\cite{torres02}). These corrections may be needed to remove small radial velocity measurement errors due to line blending and/or the shifting of spectral features in or out of the analysed wavelength range due to orbital motion. The corrections are found by simulating observed spectra with known radial velocities and analysing them in the same way as in the target spectra.

A set of simulated spectra were generated for each spectroscopic orbit, using HR\,1855 (broadened) as a template, for the observed phases and calculated velocities. A light ratio of 0.9 was used and no additional simulated observational noise was added. We have derived these corrections for all four of our velocity analysis methods because there was {\it a priori} no reason to assume that the corrections would be negligible for any one of them. The derived corrections were applied to the radial velocities from the target spectra and new spectroscopic
orbits were calculated.

\begin{figure} \resizebox{\hsize}{!}{\includegraphics{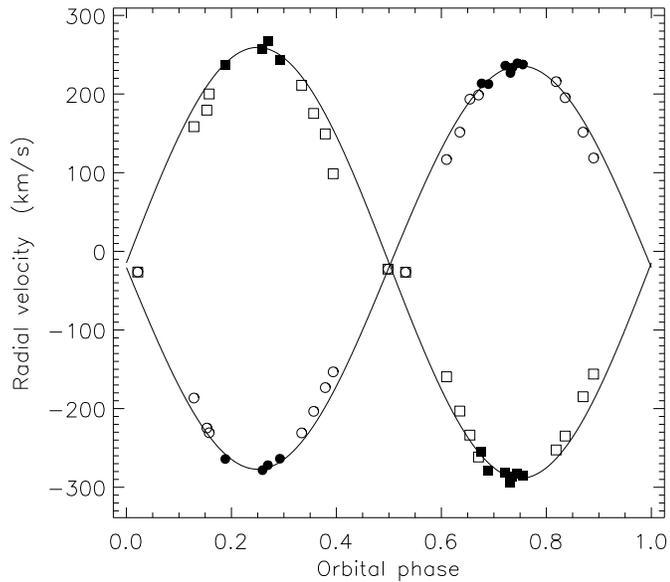}}
\caption{Radial velocities and best-fitting spectroscopic orbit measured
from analysing the $\lambda$4471 spectral line with {\sc todcor}. No
velocity corrections have been made. Circles represent velocities for
star A and squares for star B. Rejected observations are shown using open
symbols. They were rejected due to the appearance of their CCF, not the
value of the derived velocities. See text for discussion.}
\label{fig:rvorbit:uncor} \end{figure}

\begin{figure} \resizebox{\hsize}{!}{\includegraphics{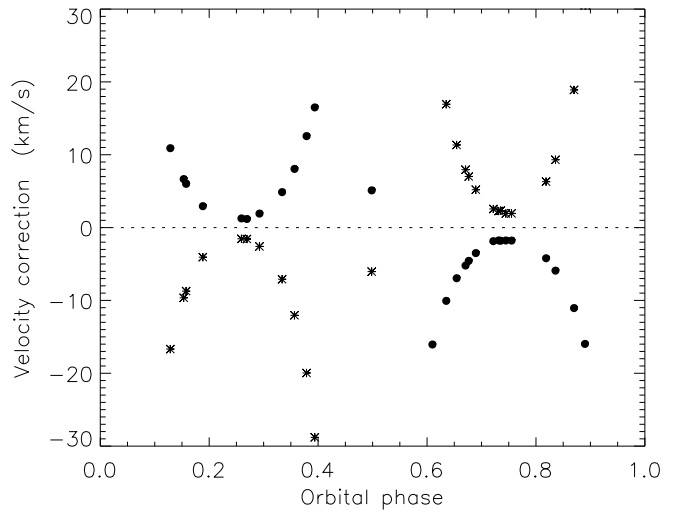}}
\caption{Corrections derived for the radial velocities measured from
the $\lambda$4471 line using {\sc todcor}. Values for star A are shown
using filled circles and those for star B using asterisks. These
corrections are an excellent indicator of the effects of line blending.}
\label{fig:rvcor} \end{figure}

\begin{figure} \resizebox{\hsize}{!}{\includegraphics{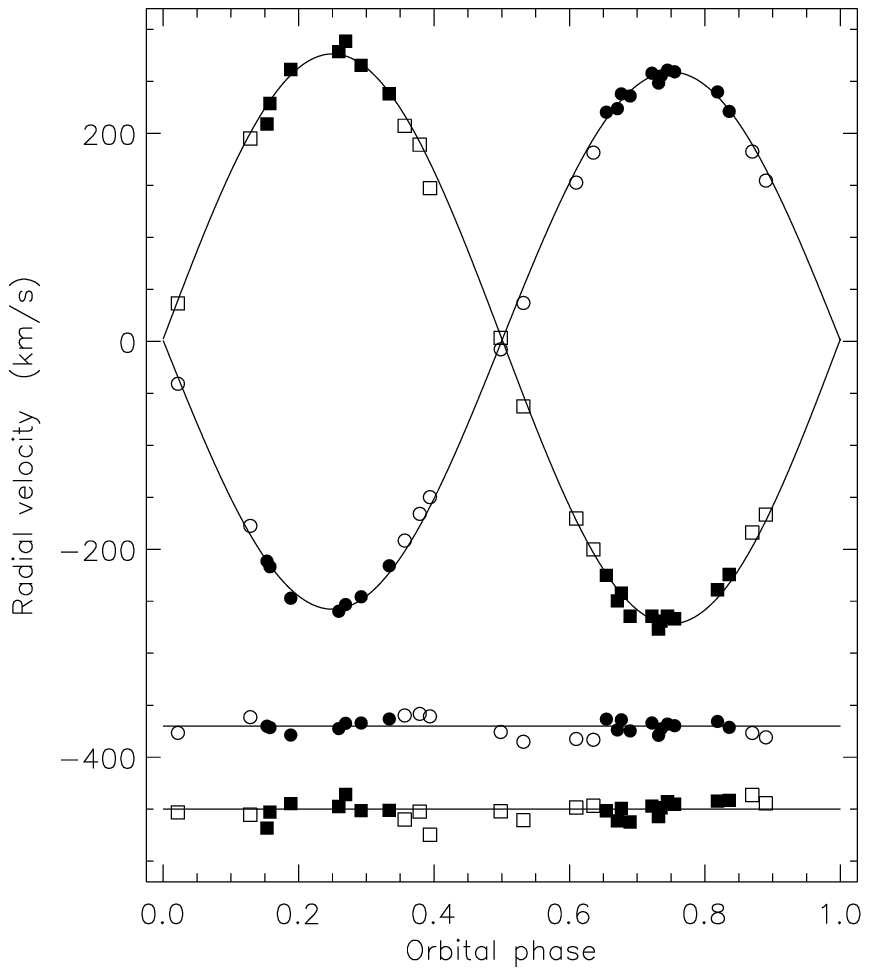}}
\caption{Radial velocities and best-fitting spectroscopic orbit measured
from a {\sc todcor} analysis of the $\lambda$4471 spectral line with with
velocity corrections subtracted. Symbols are as in Fig.~\ref{fig:rvorbit:uncor}.
The residuals of the fit are shown towards the bottom of the plot. Points
were rejected on the basis of the apearance of their CCF, but taking into
account the effects of applying the corrections.}\label{fig:rvorbit:cor}
\end{figure}

The spectroscopic orbits without velocity corrections are given in Table~\ref{table:rvorbits} and the corrected orbits are given in Table~\ref{table:rvorbits:cor}; it can be seen that in the case of DW\,Car the velocity corrections generally amount to a compensation for the systematic error caused by line blending. We have chosen the {\sc todcor} analysis of the He\,I $\lambda$4471 line to illustrate these results and the nature of the velocity corrections. The uncorrected radial velocities and best-fitting spectroscopic orbits are shown in Fig.~\ref{fig:rvorbit:uncor}, and the derived corrections are plotted in Fig.~\ref{fig:rvcor}. The corrected radial velocities are shown in Fig.~\ref{fig:rvorbit:cor}; note that the use of corrections allows more velocities to be retained as reliable, improving the quality of the resulting spectroscopic orbits.

The two cross-correlation techniques require the largest velocity corrections, which is understandable because CCFs are effectively smoothed by the rotational broadening both of DW\,Car and of the template spectrum, which was broadened to provide the closest match to the spectra of the components of DW\,Car. Whilst {\sc todcor} was introduced to minimise the effects of spectral line blending (Zucker \& Mazeh \cite{zucker94}), the two-dimensional CCF produced by the {\sc todcor} algorithm is calculated from the normal CCFs between the target and each template spectrum and between the two template spectra themselves. As the one-dimensional CCFs suffer from line blending, it is not surprising that this effect is still present in the {\sc todcor} CCFs.

The velocity corrections required by {\sc 2gauss} were a lot smaller than those for {\sc onecor} or {\sc todcor}. This is understandable because fitting a {\it double} Gaussian function explicitly compensates for spectral line blending. The velocity corrections arise mainly due to slight departures in the shape of spectral lines from Gaussians. However, the corrections of {\sc 2gauss} radial velocities have a larger effect on the resulting systemic velocities than those for {\sc onecor} and {\sc todcor}. This is again due to slightly non-Gaussian spectral line shapes, and is particularly noticeable for the $\lambda$4471 line which has a blended forbidden component on its short-wavelength side.

Velocity corrections to spectroscopic orbits calculated by disentangling were negligible, confirming in this case that the technique is an excellent way to measure spectroscopic orbits. This suggests that spectral disentangling is not significantly affected by line blending. This conclusion was also reached by Hilditch (\cite{hilditch05}).

\subsection{Final Keplerian orbit}                            \label{sec:finalorbitish}

\begin{table*} \caption{Spectroscopic orbits calculated for each component of DW\,Car,
given for each of the considered He\,I lines and for each of the four radial velocity
measurement techniques. $K_{\rm A}$ and $K_{\rm B}$ refer to the velocity semiamplitudes
of the stars and $V_{\rm \gamma,A}$ and $V_{\rm \gamma,B}$ to the systemic velocities
(all quantities in km\,s$^{-1}$). Systemic velocities cannot be measured as directly
using spectral disentangling, and no error estimates are available (see text for details).}
\label{table:rvorbits} \centering
\begin{tabular}{l l c r@{\,$\pm$\,}l r@{\,$\pm$\,}l r@{\,$\pm$\,}l c} \hline \hline
Spectral line (\AA)& \'Echelle& &\mc{Gaussian fitting}&  \multicolumn{4}{c}{Cross-correlation}  &Spectral     \\
                   & order    & &\mc{} &\mc{One-dimensional}&\mc{Two-dimensional}&disentangling\\ \hline
He I              & 55  & $K_{\rm A}$        & 259.0 & 2.2 & 252.8 & 1.5 & 254.5 & 1.6 & 263 \\
4026.189          &     & $K_{\rm B}$        & 277.8 & 1.5 & 265.5 & 3.2 & 267.3 & 3.3 & 277 \\
                  &     & $V_{\rm \gamma,A}$ & $-$15.2 & 2.0 & $-$0.4 & 1.5 & $-$1.3 & 1.6 &     \\
                  &     & $V_{\rm \gamma,B}$ & $-$14.5 & 1.3 & $-$0.7 & 3.1 & $-$2.3 & 3.2 &     \\[2pt]
He I              & 51  & $K_{\rm A}$        & 261.4 & 1.3 & 259.8 & 1.8 & 257.2 & 1.1 & 263 \\
4387.928          &     & $K_{\rm B}$        & 280.6 & 2.0 & 280.2 & 2.2 & 274.8 & 2.0 & 279 \\
                  &     & $V_{\rm \gamma,A}$ &  $-$3.4 & 0.9 & $-$1.3 & 1.7 & $-$1.0 & 1.0 &     \\
                  &     & $V_{\rm \gamma,B}$ &  $-$3.8 & 1.5 & $-$1.0 & 2.2 & $-$0.2 & 1.9 &     \\[2pt]
He I              & 50  & $K_{\rm A}$        & 261.9 & 1.5 & 252.2 & 2.0 & 255.4 & 1.5 & 258 \\
4471.681          &     & $K_{\rm B}$        & 281.5 & 2.2 & 274.6 & 3.1 & 273.7 & 2.7 & 277 \\
                  &     & $V_{\rm \gamma,A}$ & $-$30.9 & 1.4 & $+$9.9 & 2.0 & $+$0.2 & 1.5 &     \\
                  &     & $V_{\rm \gamma,B}$ & $-$23.7 & 2.0 & $+$8.3 & 3.1 & $+$.55 & 2.7 &     \\[2pt]
He I              & 45  & $K_{\rm A}$        & 254.1 & 2.0 & 251.6 & 4.0 & 251.3 & 3.6 & 255 \\
4921.929          &     & $K_{\rm B}$        & 276.3 & 1.1 & 269.1 & 3.0 & 270.6 & 2.7 & 275 \\
                  &     & $V_{\rm \gamma,A}$ & $-$11.3 & 1.8 & $-$3.4 & 4.0 & $-$4.4 & 3.4 &     \\
                  &     & $V_{\rm \gamma,B}$ & $-$13.0 & 1.0 & $-$9.9 & 3.0 & $-$9.8 & 2.6 &     \\
He I              & 33  & $K_{\rm A}$        &  \multicolumn{6}{c}{\ }                 & 263 \\
6678.149          &     & $K_{\rm B}$        &  \multicolumn{6}{c}{\ }                 & 281 \\
\hline \end{tabular} \end{table*}

\begin{table*} \caption{As Table~\ref{table:rvorbits} but with velocity corrections applied (see text for details).}
\label{table:rvorbits:cor} \centering
\begin{tabular}{l l c r@{\,$\pm$\,}l r@{\,$\pm$\,}l r@{\,$\pm$\,}l c} \hline \hline
Spectral line (\AA)&\'Echelle& &\mc{Gaussian fitting}&   \multicolumn{4}{c}{Cross$-$correlation}   & Spectral    \\
                   &order    & &\mc{} &\mc{One$-$dimensional}&\mc{Two$-$dimensional}& disentangling\\ \hline
He I              & 55  & $K_{\rm A}$        & 263.1 & 2.1 & 262.8 & 1.1 & 259.3 & 1.6 & 263 \\
4026.189          &     & $K_{\rm B}$        & 282.2 & 1.5 & 275.2 & 2.0 & 275.0 & 1.8 & 277 \\
                  &     & $V_{\rm \gamma,A}$ &$-$0.4 & 1.9 &   3.6 & 1.1 &   2.0 & 1.5 &     \\
                  &     & $V_{\rm \gamma,B}$ &   0.4 & 1.3 &   2.5 & 1.9 &   0.8 & 1.6 &     \\[2pt]
He I              & 51  & $K_{\rm A}$        & 254.6 & 1.0 & 251.4 & 1.2 & 252.4 & 1.1 & 263 \\
4387.928          &     & $K_{\rm B}$        & 272.1 & 1.7 & 266.1 & 2.0 & 267.6 & 1.8 & 279 \\
                  &     & $V_{\rm \gamma,A}$ & $-$17.5 & 0.9 & $-$18.9 & 1.8 & $-$18.6 & 1.0 &     \\
                  &     & $V_{\rm \gamma,B}$ & $-$15.5 & 1.5 & $-$19.2 & 1.1 & $-$18.1 & 1.6 &     \\[2pt]
He I              & 50  & $K_{\rm A}$        & 262.0 & 1.2 & 257.6 & 1.6 & 257.8 & 1.4 & 258 \\
4471.681          &     & $K_{\rm B}$        & 281.6 & 1.8 & 275.9 & 2.6 & 273.9 & 2.2 & 277 \\
                  &     & $V_{\rm \gamma,A}$ &   0.0 & 1.1 &   5.7 & 2.6 &   1.3 & 1.3 &     \\
                  &     & $V_{\rm \gamma,B}$ &   3.0 & 1.6 &   3.0 & 1.6 &   1.4 & 2.0 &     \\[2pt]
He I              & 45  & $K_{\rm A}$        & 255.0 & 2.0 & 252.4 & 2.8 & 252.5 & 2.1 & 256 \\
4921.929          &     & $K_{\rm B}$        & 276.9 & 1.1 & 274.1 & 1.9 & 272.2 & 2.0 & 275 \\
                  &     & $V_{\rm \gamma,A}$ &  $-$6.2 & 1.5 &  $-$7.4 & 2.6 &  $-$6.9 & 1.9 &     \\
                  &     & $V_{\rm \gamma,B}$ &  $-$8.4 & 1.1 &  $-$9.2 & 1.7 & $-$10.5 & 1.8 &     \\
He I              & 33  & $K_{\rm A}$        &  \multicolumn{6}{c}{\ }                 & 263 \\
6678.149          &     & $K_{\rm B}$        &  \multicolumn{6}{c}{\ }                 & 281 \\
\hline \end{tabular} \end{table*}

The final velocity semiamplitudes for each radial velocity measurement technique are given in Table~\ref{table:finalorbit} along with our adopted values (velocity semiamplitudes from spectral disentangling and systemic velocities from {\sc 2gauss}). The velocity semiamplitudes are in good agreement with the values derived by Ferrer et al.\ (\cite{ferrer85}): $K_{\rm A} = 261 \pm 10$\kms\ and $K_{\rm B} = 278 \pm 10$\kms. Our systemic velocities are in acceptable agreement with the $-10 \pm 5$ and $-16 \pm 5$\kms\ found by Ferrer et al.

We have shown that velocity corrections are needed for {\sc onecor} and {\sc 2gauss} as well as for {\sc todcor}, and that both cross-correlation techniques seem to suffer from line blending even when these corrections have been included. We have also discovered that small differences in how the CCFs are calculated (including the size of the spectral range and whether the lowest frequency components of the Fourier transforms are suppressed) can make a significant difference to the uncorrected velocity semiamplitudes. These differences normally vanish when the velocity corrections are applied, illustrating how important the corrections can be. However, strongly blended lines continue to give unreliable velocities even when corrections are applied. Calculating corrections iteratively, using already corrected orbits, improves this slightly, but cross-correlation cannot reliably deal with spectra which are strongly affected by line blending.

Averaged spectroscopic orbits for each analysis technique have been calculated by finding the mean and the standard error of the velocity-corrected orbital parameters in Table~\ref{table:rvorbits:cor} for different spectral lines (Table~\ref{table:finalorbit}). It can be seen that the velocity semiamplitudes calculated using {\sc onecor} and {\sc todcor} are lower than those from {\sc 2gauss} and disentangling. The main problem in determining these spectroscopic orbits is line blending, which will always act to reduce $K_{\rm A}$ and $K_{\rm B}$. Therefore the techniques which are least affected by line blending should give higher $K_{\rm A}$ and $K_{\rm B}$, implying that disentangling is best, closely followed by {\sc 2gauss}, whereas the cross-correlation techniques, {\sc onecor} and {\sc todcor}, are significantly affected.

We have adopted the averaged spectroscopic orbit (and standard error) from disentangling as the best result, with the systemic velocities from {\sc 2gauss} (note that this is not the final orbit: see below). The good agreement between this and the average orbit for {\sc 2gauss} shows that the technique of fitting double Gaussians also performs very well in situations such as this. It is important to note that the results we derived using spectral disentangling can only be fully trusted because the results from other techniques confirm that we have chosen the correct peaks in the surfaces of fit quality versus $K_{\rm A}$ and $K_{\rm B}$ (Fig.~\ref{fig:k1k2grid}), although in every case this peak was a global maximum. We have tabulated one set of radial velocities for the interested reader (Table~\ref{table:rvs}). We chose the {\sc 2gauss} analysis of the He\,I $\lambda$4471 line, as {\sc 2gauss} seems to be more reliable than {\sc onecor} or {\sc todcor}, and the $\lambda$4471 line velocities give a similar result to the adopted final orbit.

It is also noticeable from Table~\ref{table:finalorbit} that the agreement between different analyses of the same spectral line is greater than that between the same analysis method used on different lines. This suggests that the systematic errors of different techniques are smaller than the random errors, although it must be remembered that the systematic errors may remain at full strength after the results from several lines have been averaged. The systemic velocities from different \'echelle orders have a poorer agreement than expected; any further investigation of this will require the use of spectra from a dEB which exhibits a far richer set of spectral lines.

\begin{table} \caption{Keplerian spectroscopic orbital parameters for DW\,Car. The mean and
standard error are given for each of the four radial velocity analysis techniques, and for
the final results. All quantities are in units of \kms).}
\label{table:finalorbit} \centering
\begin{tabular}{l r@{\,$\pm$\,}l r@{\,$\pm$\,}l r@{\,$\pm$\,}l r@{\,$\pm$\,}l} \hline \hline
       & \mc{$K_{\rm A}$} & \mc{$K_{\rm B}$} & \mc{${\Vsys}_A$} & \mc{${\Vsys}_B$} \\ \hline
{\sc 2gauss} & 258.7 & 1.9 & 278.2 & 2.0 & $-$7.3 & 3.7 & $-$6.3 & 4.0 \\
{\sc onecor} & 256.0 & 2.3 & 272.8 & 2.0 & $-$4.3 & 4.9 & $-$5.7 & 4.6 \\
{\sc todcor} & 255.5 & 1.5 & 272.2 & 1.4 & $-$5.6 & 4.2 & $-$6.6 & 4.1 \\
Spec.\,dis.  & 260.4 & 1.5 & 277.8 & 0.9 & \multicolumn{4}{c}{}  \\ \hline
Adopted        & 260.4 & 1.5 & 277.8 & 0.9 & $-$7.3 & 3.7 & $-$6.3 & 4.0 \\
\hline \end{tabular} \end{table}

\begin{table} \caption{Corrected radial velocities from a {\sc 2gauss} analysis
of the He\,I $\lambda$4471 spectral line. Note that these data are only a small
part of those used to calculate the final spectroscopic orbits. The columns marked
$O-C$ give the observed minus calculated velocity residuals.}
\label{table:rvs} \centering
\begin{tabular}{lrrrrr} \hline \hline
HJD $-$     & Weight & \mc{Radial velocities}   & \mc{$O-C$}     \\
2\,400\,000 &        &    star A   &   star B   & \multicolumn{1}{c}{A} &
                                                  \multicolumn{1}{c}{B} \\ \hline
51144.83060 &  1.0   &    257.4    &   -279.8   &  -3.0  &  -2.0 \\
51145.84399 &  0.0   &     25.0    &     13.2   &  28.5  &   8.8 \\
51146.86755 &  1.0   &   -256.0    &    284.5   &   5.2  &   4.5 \\
51147.86649 &  0.0   &     -1.4    &     48.9   &  34.9  &   9.4 \\
51148.80902 &  1.0   &    247.7    &   -278.5   & -12.1  &  -1.4 \\
51149.87170 &  0.0   &    106.7    &     66.2   &  55.4  & 120.4 \\
51150.83746 &  1.0   &   -262.2    &    276.5   &   0.5  &  -5.1 \\
51152.73657 &  1.0   &    240.0    &   -275.1   &  -2.8  & -16.1 \\
51172.69544 &  2.0   &    256.2    &   -273.6   &  -1.2  &   1.0 \\
51174.78088 &  2.0   &   -252.7    &    269.4   &   1.3  &  -2.8 \\
51176.80734 &  1.0   &    246.9    &   -245.4   &   9.3  &   8.0 \\
51178.84953 &  1.0   &   -208.1    &    226.8   &  -1.6  &   5.4 \\
51179.87450 &  0.0   &   -192.6    &    216.3   &  -2.2  &  12.0 \\
51180.81349 &  1.0   &    230.1    &   -238.3   &   5.8  &   1.0 \\
51181.87490 &  1.0   &    193.7    &   -219.9   &  -2.4  & -10.8 \\
51182.86220 &  0.0   &   -184.0    &    209.9   &  -2.2  &  14.8 \\
51184.86864 &  0.0   &    167.6    &   -193.0   &   1.1  & -15.6 \\
51187.87364 &  1.0   &   -220.4    &    233.8   &  -4.4  &   2.1 \\
51189.86682 &  1.0   &    215.1    &   -223.6   &  -0.5  &   6.2 \\
51190.84867 &  0.0   &   -207.3    &    118.5   & -44.1  & -56.6 \\
51191.86279 &  1.0   &   -229.4    &    238.5   &  -9.3  &   2.5 \\
51192.80846 &  1.0   &    199.0    &   -201.8   &   8.4  &   1.4 \\
51193.87953 &  1.0   &    232.0    &   -244.4   &  -2.1  &   5.2 \\
51197.85496 &  1.0   &    230.3    &   -254.2   &   0.7  &  -9.4 \\
51199.87004 &  2.0   &   -247.1    &    254.8   &  -3.5  &  -6.4 \\
51209.72395 &  1.0   &    164.3    &   -196.2   &  -1.9  & -19.1 \\
51382.51044 &  2.0   &    263.2    &   -270.3   &   1.9  &   8.6 \\
51386.50761 &  2.0   &    261.8    &   -273.0   &   0.5  &   5.9 \\
51563.86655 &  2.0   &   -221.5    &    251.4   &   6.1  &   7.4 \\
\hline \end{tabular} \end{table}

\subsection{Final spectroscopic orbit}                                  \label{sec:finalorbit}

\begin{table} \caption{Final spectroscopic orbit after correction for non-Keplerian
effects. $i$ and $a$ are the orbital inclination and semimajor axis, respectively.}
\label{table:finalorbit2} \centering
\begin{tabular}{lr@{\,$\pm$\,}lr@{\,$\pm$\,}l} \hline \hline
                        &    \mc{Star A}     &    \mc{Star B}     \\ \hline
$K$ (Keplerian) (\kms)  &    260.4 & 1.5     &    277.8 & 0.9     \\
$K$ (corrected) (\kms)  &    261.7 & 1.6     &    279.3 & 1.2     \\
Mass ratio              & \multicolumn{4}{c}{$0.9370 \pm 0.0070$} \\
$a \sin i$ (\Rsun)      & \multicolumn{4}{c}{$14.192 \pm 0.052$}  \\
$M \sin^3 i$ (\Msun)    &    11.25 & 0.12    &    10.54 & 0.14    \\
\hline \end{tabular} \end{table}

The spectroscopic orbits fitted so far have been purely Keplerian, i.e.\ we have assumed that the velocities of the centres of light of the stars are identical to those of the their centres of mass. In close binaries, though, the effects of reflection and ellipsoidal stellar shape can cause the observed velocities to be significantly different to the centre-of-mass velocities. These effects are accurately modelled by the Wilson-Devinney code (see Section \ref{sec:lc}), which can be used to correct individual radial velocities using a given system geometry. However, our adopted velocity semiamplitudes were calculated using spectral disentangling. With this technique the spectroscopic orbit is calculated directly without the intermediate step of calculating radial velocities. We therefore cannot apply radial velocity corrections and so must calculate corrections which can be applied to the velocity semiamplitudes themselves.

We have calculated spectroscopic orbits for the {\sc 2gauss} radial velocities, using the Wilson-Devinney code both with and without the inclusion of non-Keplerian effects. This has been done for \'echelle orders 55, 51, 50 and 45. The mean value and standard error of the resulting corrections to the velocity semiamplitudes (in the sense $K_{\rm true} - K_{\rm Keplerian} = \Delta K$) are $\Delta K_{\rm A} = +1.3 \pm 0.1$\kms\ and $\Delta K_{\rm B} = +1.5 \pm 0.2$\kms. This indicates that in this case fitting Keplerian orbits caused the velocity amplitudes (and so the masses) to be systematically underestimated by a small but significant amount when fitting a Keplerian orbit. This is the expected result, because the light centres for each star are shifted towards the companion due to their mutual irradiation, causing the measured radial velocities to be underestimates of the motion of the centre of mass.

The final spectroscopic orbit, with the non-Keplerian correction applied, is given in Table~\ref{table:finalorbit2}.
The resulting minimum masses are $M_{\rm A} \sin^3 i = 11.25 \pm 0.12$\Msun\ and $M_{\rm B} \sin^3 i = 10.54 \pm 0.14$\Msun\ where $i$ is the orbital inclination. The corrections to the systemic velocities are negligible and have not been applied.

\subsection{Spectroscopic light ratio}                                        \label{sec:spec:lr}

The light ratio of the two stars is poorly defined by the light curves of the system (see below) so we have derived a spectroscopic light ratio to help constrain the light curve solutions. The {\sc todcor} algorithm was used to analytically determine the light ratio for each spectral line analysed (see Sect.~\ref{sec:rv:todcor} and Table~\ref{table:specdata}), yielding an average value and standard deviation of $\frac{\ell_{\rm B}}{\ell_{\rm A}} = 0.873 \pm 0.042$. The wavelength dependence of the light ratio is negligible between the Balmer jump and 5500\,\AA\ (see Section~\ref{sec:lc:uvby}). This is a reliable equivalent width ratio measurement because the components of DW\,Car have very similar spectral characteristics, and the same template spectrum was used for both stars.

The equivalent widths of the He\,I $\lambda$4026 and $\lambda$4471 lines for both stars were also measured directly using several spectra with large velocity separations and high signal to noise ratios. The component lines are still slightly blended so we have used the deblending feature of the {\sc iraf}\footnote{{\sc iraf} is distributed by the National Optical Astronomical Observatories, which are operated by the Association of Universities for Research in Astronomy, Inc.\ under contract with the National Science Foundation.} task {\sc splot} to measure the line strengths. The mean values and standard deviations are $0.87 \pm 0.03$ and $0.89 \pm 0.03$ for the $\lambda$4026 and $\lambda$4471 lines respectively, in good agreement with the {\sc todcor} results.

The He\,I lines are intrinsically slightly stronger in star B ($\Teff = 26\,500$\,K) than in star A ($\Teff = 27\,900$\,K), so the equivalent width ratios are slightly different to light ratios. A correction was derived using intrinsic equivalent width ratios taken from {\sc uclsyn} and was found to be $0.97 \pm 0.01$. The final light ratio, using the equivalent width measurements, is $0.84 \pm 0.04$ for the $\lambda$4026 line, which is close to the central wavelength of the Str\"omgren {\it v} passband, 4110\,\AA\ (Str\"omgren \cite{stromgren63}).

\subsection{Spectroscopic helium abundance}                             \label{sec:spec:He}

\begin{figure*} \resizebox{\hsize}{!}{\includegraphics{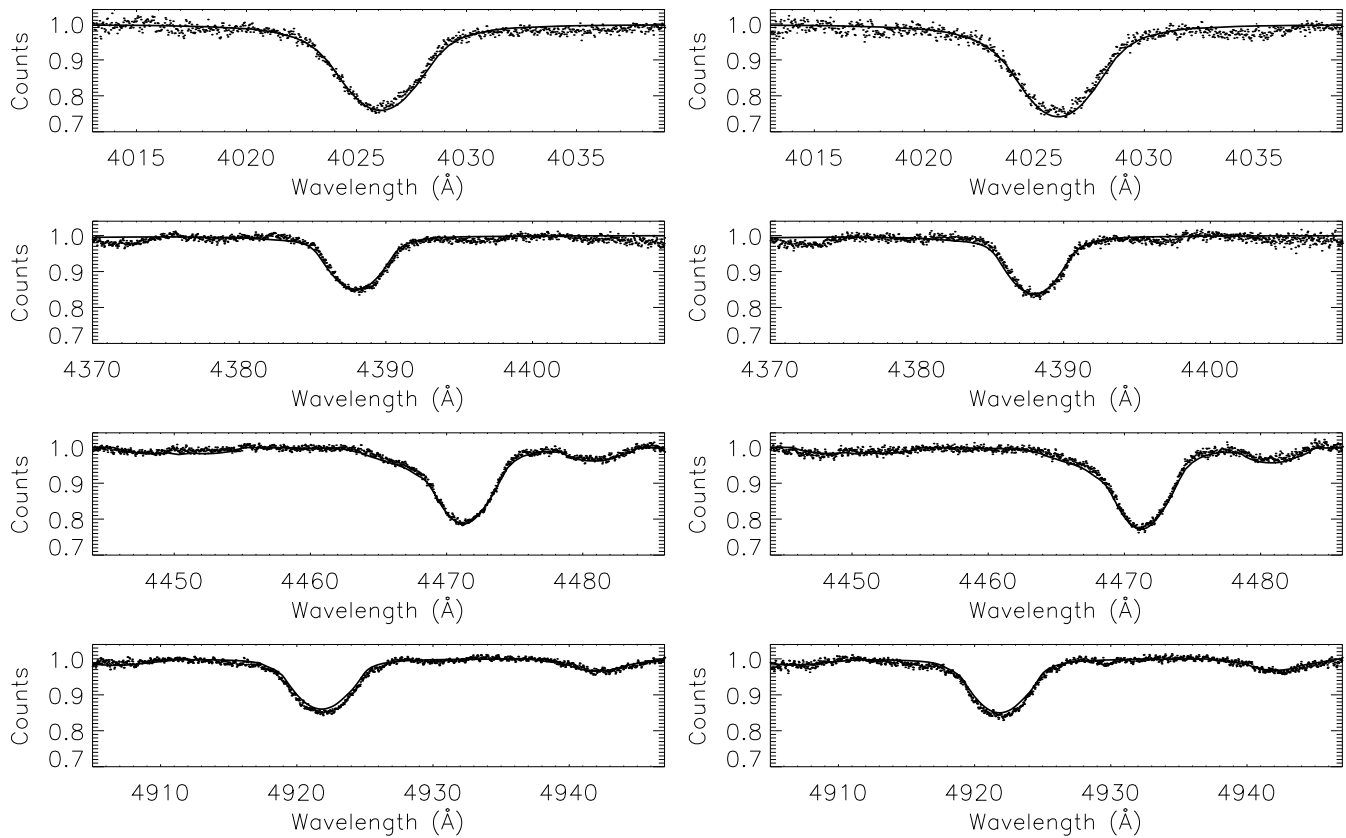}}
\caption{Plot of the normalised disentangled profiles of four helium
lines for each star compared to theoretical spectra calculated for
helium abundance $Y = 0.28$ and the effective temperatures and light
ratio found in our analysis. The left-hand panels are for star A and
the right-hand panels for star B. The observed spectra are plotted
using points and the theoretical spectra using solid lines.}
\label{fig:spec:he} \end{figure*}

At the suggestion of the referee we have attempted to derive surface helium abundances for the components of DW\,Car. The high rotational velocities mean that this is difficult, but spectroscopic abundance analyses have been successfully performed on similar objects by Pavlovski \& Hensberge (\cite{pav05}) and Pavlovski et al.\ (\cite{pav06}). Our data contain insufficient information to significantly constrain the metal abundances of the components of DW\,Car.

Model atmospheres were calculated using the {\sc atlas9} code (Kurucz \cite{kurucz79}). Non-LTE atomic level populations were obtained for hydrogen and helium using the {\sc detail} code and synthetic spectra were constructed using the {\sc surface} code (Butler \& Giddings \cite{butler85}). We adopted the effective temperatures and light ratio for DW\,Car found below and calculated the residuals of the best fit for several different helium abundances.

Formally, the best fits are found for a helium abundance equal to or slightly lower than the solar value (Fig.~\ref{fig:spec:he}), but with a large uncertainty. Varying the metal abundance within reasonable limits has no significant effect. However, whilst the fit of the lines at 4388\,\AA\ and 4471\,\AA\ are excellent, the 4026\,\AA\ line is too shallow in the synthetic spectrum whilst the 4922\,\AA\ line is too deep. The investigation of these effects is beyond the scope of this work. We conclude that the spectra are consistent with a solar or subsolar helium abundance but not with an abundance signficantly greater than solar.

The rotational velocities were also optimised when fitting the theoretical spectra to the disentangled spectra, and the best-fitting values are $V_{\rm A} \sin i = 182 \pm 3$\kms\ and $V_{\rm B} \sin i = 177 \pm 3$\kms.


\section{Light curve analysis}                                                                  \label{sec:lc}

\begin{figure*} \resizebox{\hsize}{!}{\includegraphics{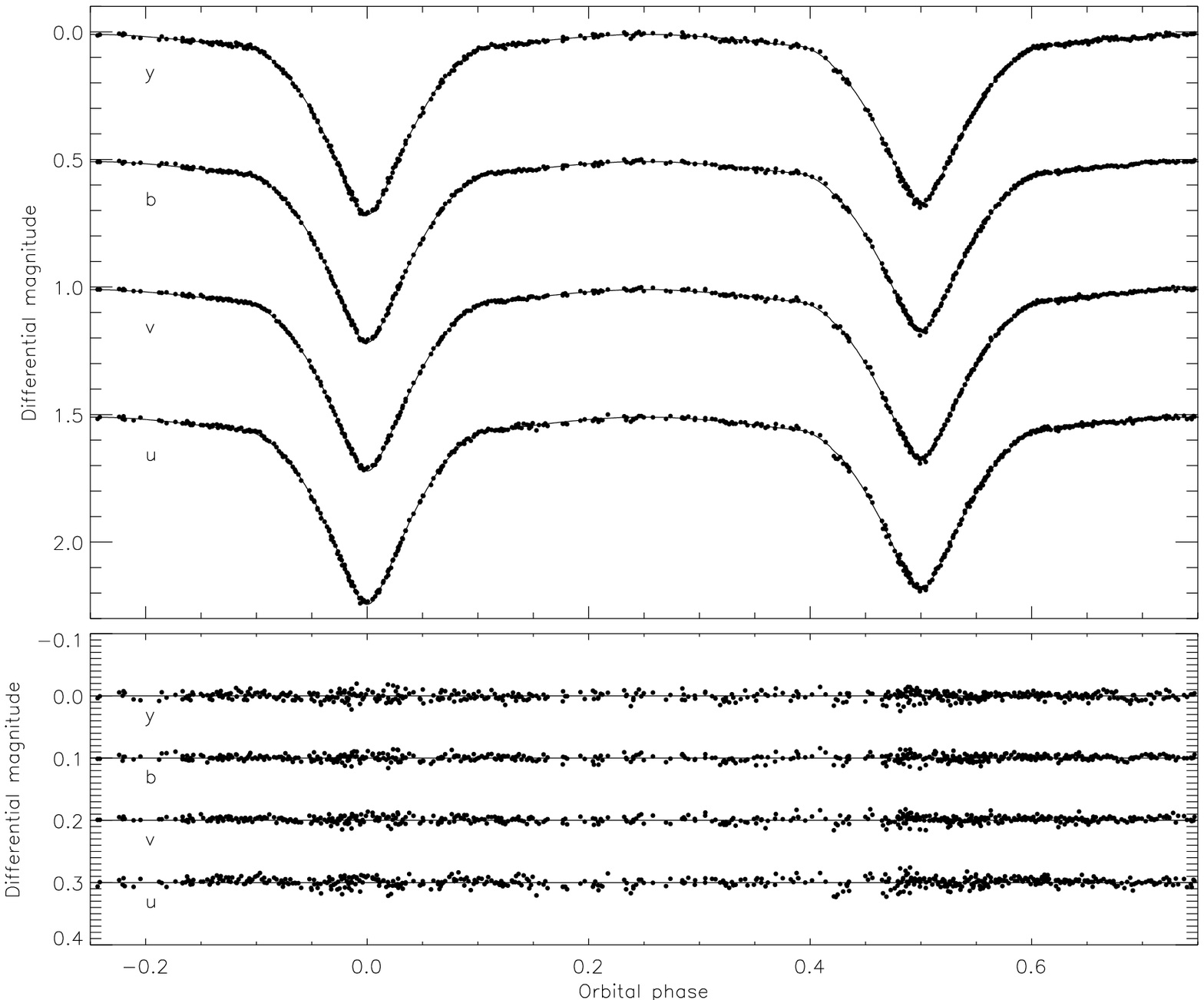}}
\caption{Light curves and best {\sc wd2003} fit with the potential of
star A fixed at 4.1706 (upper panel) and the residuals of the fit
(lower panel) shown with an expanded scale in magnitudes.}
\label{fig:lc:lc} \end{figure*}

\begin{table*}
\caption{{\sc wd2003} light curve solutions for DW\,Car. A light ratio was used to constrain the
solution for the $v$ light curve, and the resulting potential of star A was fixed for the other
light curves (see text for further details). The quoted errors (except for the adopted values in
the final column) are the standard errors calculated by {\sc wd2003}.}
\label{table:lc:lc} \centering
\begin{tabular}{l c r@{\,$\pm$\,}l r@{\,$\pm$\,}l r@{\,$\pm$\,}l r@{\,$\pm$\,}l r} \hline \hline
Parameter        & Potential &   \mc{$v$}      &    \mc{$y$}     &   \mc{$b$}      &   \mc{$u$}      & Adopted \\
                 & of star A &        &        &        &        &        &        &        &        & values  \\
\hline
Light ratio          & 4.101 &   \mc{0.800}    &  0.796 & 0.012  &  0.797 & 0.009  &  0.769 & 0.034  &\\
                     & 4.132 &   \mc{0.840}    &  0.840 & 0.014  &  0.840 & 0.010  &  0.813 & 0.034  &\\
                     & 4.157 &   \mc{0.880}    &  0.879 & 0.013  &  0.881 & 0.010  &  0.858 & 0.032  &\\[3pt]
Potential of         & 4.101 &  4.225 & 0.007  &  4.236 & 0.009  &  4.228 & 0.007  &  4.248 & 0.011  &\\
star B               & 4.132 &  4.192 & 0.007  &  4.200 & 0.009  &  4.193 & 0.007  &  4.211 & 0.010  &\\
                     & 4.157 &  4.157 & 0.007  &  4.167 & 0.008  &  4.158 & 0.007  &  4.172 & 0.009  &\\[3pt]
Fractional radius    & 4.101 &   \mc{0.3237}   &   \mc{0.3236}   &   \mc{0.3237}   &   \mc{0.3237}   &      0.3203\\
of star A            & 4.132 &   \mc{0.3203}   &   \mc{0.3203}   &   \mc{0.3203}   &   \mc{0.3203}   &$\pm$ 0.0030\\
                     & 4.157 &   \mc{0.3176}   &   \mc{0.3175}   &   \mc{0.3175}   &   \mc{0.3175}   &\\[3pt]
Fractional radius    & 4.101 & 0.2993 & 0.0007 & 0.2982 & 0.0009 & 0.2990 & 0.0007 & 0.2970 & 0.0011 &      0.3019\\
of star B            & 4.132 & 0.3026 & 0.0007 & 0.3018 & 0.0009 & 0.3025 & 0.0007 & 0.3007 & 0.0010 &$\pm$ 0.0037\\
                     & 4.157 & 0.3063 & 0.0007 & 0.3052 & 0.0009 & 0.3061 & 0.0007 & 0.3048 & 0.0010 &\\[3pt]
Orbital inclination  & 4.101 &  85.76 & 0.07   &  86.00 & 0.09   &  85.96 & 0.07   &  86.13 & 0.08   &     85.72\\
(degrees)            & 4.132 &  85.61 & 0.07   &  85.77 & 0.04   &  85.75 & 0.04   &  85.74 & 0.06   &$\pm$ 0.28\\
                     & 4.157 &  85.63 & 0.07   &  85.73 & 0.08   &  85.70 & 0.07   &  85.83 & 0.09   &\\[3pt]
Phase shift of primary&4.101 &   0.06 & 0.04   &  -0.02 & 0.03   &  -0.02 & 0.05   &   0.03 & 0.05   &\\
midminimum ($10^{-3}$)&4.132 &   0.08 & 0.05   &  -0.03 & 0.07   &  -0.01 & 0.05   &   0.01 & 0.05   &\\
                     & 4.157 &   0.06 & 0.04   &  -0.04 & 0.04   &   0.00 & 0.05   &   0.04 & 0.04   &\\[3pt]
Linear LD coefficient& 4.101 & -0.124 & 0.031  & -0.132 & 0.037  & -0.156 & 0.029  & -0.056 & 0.038  &\\
of star A (using the & 4.132 & -0.084 & 0.031  & -0.091 & 0.038  & -0.113 & 0.030  & -0.023 & 0.038  &\\
square-root law)     & 4.157 & -0.125 & 0.036  & -0.114 & 0.043  & -0.137 & 0.036  & -0.039 & 0.045  &\\[3pt]
Linear LD coefficient& 4.101 & -0.093 & 0.042  & -0.087 & 0.050  & -0.117 & 0.041  & -0.069 & 0.058  &\\
of star B (using the & 4.132 & -0.095 & 0.042  & -0.068 & 0.052  & -0.109 & 0.042  & -0.017 & 0.057  &\\
square-root law)     & 4.157 & -0.145 & 0.043  & -0.114 & 0.052  & -0.149 & 0.042  & -0.065 & 0.058  &\\[3pt]
Size of residuals              & 4.101 & \mc{5.21} & \mc{6.20} & \mc{4.97} & \mc{7.15} \\
$\sigma_{\rm rms}$ ($m$mag)    & 4.132 & \mc{5.16} & \mc{6.23} & \mc{5.00} & \mc{7.04} \\
                               & 4.157 & \mc{5.39} & \mc{6.23} & \mc{5.02} & \mc{7.04} \\
\hline\end{tabular}\end{table*}


The light curves of DW\,Car contain 518 photoelectric observations in each of the $uvby$ passbands (Paper\,I). The two eclipses are relatively long in phase units and are both about 0.7\,mag deep, indicating that the system is composed of two similar and close stars. We have used the 2003 version (Van Hamme \& Wilson \cite{vanhamme03}) of the Wilson-Devinney code (Wilson \& Devinney \cite{wilson71}; Wilson \cite{wilson79}, \cite{wilson93}), hereafter called {\sc wd2003}, to model the light curves, as Roche geometry is needed to adequately represent the tidal deformation of the two stars. {\sc wd2003} optionally uses the predictions of the {\sc atlas9} model atmospheres (Kurucz \cite{kurucz93}) to connect the flux ratios of the two stars in different passbands.

Initial investigations showed that the orbital eccentricity is negligible, so we assumed that the orbit is circular. The rotational velocities of the stars were fixed at the synchronous values as these are consistent with the spectral line widths. The bolometric albedo coefficients and gravity brightening exponents were fixed at 1.0, which is the expected value for stars with radiative envelopes (Claret \cite{claret98}). The mass ratio was fixed at the spectroscopic result of 0.937 and we chose to model reflection using the faster and simpler of the two alternatives implemented in {\sc wd2003}. We have used grid sizes of $N=20$ for each star; solutions with finer grids gave negligibly different results.

We expected that there would be a small amount of contaminating `third' light in the light curves of DW\,Car, as its Be star nature indicates that there is substantial circumbinary material. Third light was therefore included as a fitting parameter in the light curve analysis, but the optimised values were all consistent with, and marginally lower than, zero. We therefore have assumed that third light is zero. A small amount of third light (up to about 5\%, which is an order of magnitude greater than the best-fitting values) will have a negligible effect on the light curve solutions.

Limb darkening coefficients were initially set to values found by bilinear interpolation in effective
temperature and surface gravity from the coefficients tabulated by Van Hamme (\cite{vanhamme93}), but in our
final solutions we have always optimised these values to avoid the theoretical dependence caused by using
coefficients calculated from stellar atmosphere models. The choice of linear, logarithmic or square-root
law has a negligible effect on the solutions so we have chosen to use the latter, as this provides the best fit to
model atmosphere predictions for hot stars (Van Hamme \cite{vanhamme93}). Fixing the limb darkening coefficients
at either the Van Hamme (\cite{vanhamme93}) or Claret (\cite{claret00}) theoretical values causes a negligible
change in the final solution -- of the order of 0.1\% in the stellar radii. The optimised values are generally in good agreement with the theoretically predicted coefficients.

The radii found by fitting the light curves with {\sc wd2003} are strongly correlated with each other and with the light ratio, resulting in one of many different light curve solutions (i.e., local minima in parameter space) being found depending on the initial fitting estimates. The residuals of the fit are approximately constant for ratios of the radii between 0.8 and 1.2. We have therefore used the light ratio of $0.84 \pm 0.04$ found in Sect.~\ref{sec:spec:lr} to constrain the solution of the $v$ band light curve. This was done by fixing the potential of star A at values which gave the desired light ratio and plus or minus its uncertainty. These values for potential were then fixed for the $uby$ light curve solutions. The potential of star B, the orbital inclination, the phase shift between primary midminimum and phase zero, and the light contributions for the two stars were optimised for each light curve. The linear terms in the square-root limb darkening law were also optimised. The effective temperatures of the two stars were fixed at nominal values (27\,500 and 26\,750\,K), but these do not affect the final solution because we are fitting for the light contributions of both stars directly. Fig.~\ref{fig:lc:lc} shows the light curve fits and residuals.

Table~\ref{table:lc:lc} contains the values of the optimised parameters, as well as the standard errors calculated by {\sc wd2003}, for each light curve and for each of the three potentials of star A (4.101, 4.132 and 4.157) which are required to reproduce the light ratios 0.80, 0.84 and 0.88 for the $v$ light curve. We have adopted the results for a star A potential of 4.132 as the final values. The uncertainties in these light curve parameter values have been calculated by adding in quadrature an uncertainty from the variation for different potentials to the mean standard error for each parameter. Additional uncertainties due to the treatment of third light and limb darkening are negligible.

\subsection{Str\"omgren indices for each star}                    \label{sec:lc:uvby}

Separate Str\"omgren photometric indices can be derived for each star from the indices of the system and the light ratios in the $uvby$ passbands. This requires a solution of the light curve with a consistent geometry, so we have derived the flux ratios and corresponding Str\"omgren indices (and uncertainties) by fitting the light curves with the potential of star A fixed at the values found above. For each fixed potential, we have calculated the Str\"omgren indices for each star from the indices of the combined system and the light ratios in the four light curves (Table~\ref{table:lc:lc}). The uncertainties in Table~\ref{table:lc:lc} have been calculated by adding in quadrature all the relevant uncertainties, including those resulting from the range of possible values of the potential of star A. They are somewhat lower than they would be if we had simply used the final light ratio errors for each light curve, because the light ratios are correlated with each other (i.e., a change in the potential of star A causes absolute changes in the light ratios significantly larger than they change relative to each other).

The $uvby$ light curves have also been solved simultaneously to derive an effective temperature difference for the two component stars. In this case, the light ratios in different passbands must be connected, which can be achieved either using Planck law or by using the predictions of theoretical model atmospheres. The Planck law is not a good option for DW\,Car because the Balmer jump causes the stars' radiative properties to be quite different from black bodies. {\sc wd2003} optionally uses predictions from Kurucz {\sc atlas9} model atmospheres, which should be reliable in this case as the effective temperatures of the two stars are very similar. Using this option, we have obtained an effective temperature difference of ${\Teff}_{\rm A} - {\Teff}_{\rm B} = 1380 \pm 420$\,K for the stars in DW\,Car, where the error has been found by adding in quadrature the uncertainty caused by varying the potential of star A as before, and the formal error of the temperature difference calculated from the covariance matrix by {\sc wd2003}. Formal errors can be underestimates of the real uncertainties of light curve parameters (Popper \cite{popper84}; Maceroni \& Rucinski \cite{maceroni97}; Southworth et al.\ \cite{SMSd}).

\begin{table}
\caption{Light ratios of DW\,Car for a consistent geometry, with the
resulting Str\"omgren indices. Note that the quoted flux ratio errors
were not used in calculating the errors of the indices of each star.}
\label{table:lc:uvby} \centering
\begin{tabular}{lrrrrr} \hline \hline
            &       &   $y$  &  $b$  &  $v$  &  $u$     \\
\hline
Flux ratios &       &  0.840 & 0.840 & 0.840 & 0.813    \\
            & $\pm$ &  0.046 & 0.044 & 0.040 & 0.056    \\
\hline \\ \hline
            &       &   $V$  & $b-y$ & $m_1$ & $c_1$    \\
\hline
System      &       &  9.675 & 0.067 & 0.046 & $-0.017$ \\
            & $\pm$ &  0.005 & 0.005 & 0.008 & 0.008    \\[3pt]
Star A      &       & 10.307 & 0.067 & 0.046 & $-0.034$ \\
            & $\pm$ &  0.026 & 0.011 & 0.016 & 0.023    \\[3pt]
Star B      &       & 10.563 & 0.067 & 0.046 & 0.004    \\
            & $\pm$ &  0.032 & 0.014 & 0.020 & 0.030    \\
\hline\end{tabular}\end{table}


\section{Absolute dimensions}                        \label{sec:absdim}

\begin{table}
\caption{The physical parameters of the component stars of DW\,Car derived
from the results of the spectroscopic and photometric analyses. $V_{\rm eq}$
denotes the observed equatorial rotational velocities of the stars.
\newline \dag\ Calculated assuming $\lsun = 3.826${$\times$}10$^{26}$\,W
and $\Mbol\sun = 4.75$ (Zombeck 1990).}
\label{table:absdim} \centering
\begin{tabular}{l r@{\,$\pm$\,}l r@{\,$\pm$\,}l} \hline \hline
Parameter                     &    \mc{Star A}     &    \mc{Star B}     \\ \hline
Mass (\msun)                  &   11.34  &  0.12   &   10.63  &  0.14   \\
Radius (\rsun)                &   4.558  &  0.045  &   4.297  &  0.055  \\
$\log (g/{\rm cm\,s^{-3}})$   &   4.175  &  0.008  &   4.198  &  0.011  \\
$V_{\rm eq}$ (\kms)           &   183    &  3      &   178    &  3      \\
\Vsync\ (\kms)                &   173.6  &  1.7    &   163.7  &  2.1    \\
$\log\Teff$ (K)               &   4.446  &  0.016  &   4.423  &  0.016  \\
$\log(L/\lsun)$\dag           &   4.055  &  0.063  &   3.915  &  0.067  \\
\Mbol\dag                     & $-$5.39  &  0.16   & $-$5.04  &  0.17   \\ \hline
\end{tabular}\end{table}

The masses, radii and surface gravities of the two components of DW\,Car have been calculated from the results of the spectroscopic and photometric analyses. Using careful error propagation, we find that the masses and radii are known to accuracies of about 1\% (Table~\ref{table:absdim}). DW\,Car therefore joins the (very short) list of well-studied high-mass unevolved dEBs, alongside QX\,Car (Andersen et al.\ \cite{andersen83}) and V578\,Mon (Hensberge, Pavlovski \& Verschueren \cite{hensberge00}).

We have derived the effective temperatures of the two stars from the Str\"omgren indices found in Section~\ref{sec:lc:uvby} using the calibration of Moon \& Dworetsky (\cite{moondw85}) contained in the code {\sc uvbybeta} (Moon \cite{moon85}). Using the individual Str\"omgren indices for each star, we find ${\Teff}_{\rm A} = 27\,800 \pm 1100$ and ${\Teff}_{\rm B} = 26\,500 \pm 1300$\,K, where the error does not include any uncertainty in the calibration and its main contribution comes from the uncertainty in the $m_1$ indices. However, the equivalent effective temperature of the system is much better defined: ${\Teff}_{\rm A+B} = 27\,230 \pm 390$\,K, and the temperature difference is also well defined from the light curve analysis: ${\Teff}_{\rm A} - {\Teff}_{\rm B} = 1380 \pm 420$\,K (formal error). Using these quantities, and including an estimated uncertainty in the photometric calibration, we find temperatures of $27\,900 \pm 1000$ and $26\,500 \pm 1000$ for the two stars. The calibrations of Davis \& Shobbrook (\cite{davis77}) and Napiwotzki, Sch\"onberner \& Wenske (\cite{napi93}) give temperatures in good agreement with these values.

The equatorial rotational velocities quoted for each star were found by comparing the disentangled spectra of the stars to theoretical spectra in order to determine the surface helium abundance (Section~\ref{sec:spec:He}). They are slightly greater than the synchronous value despite the short orbital period of DW\,Car. This indicates that it is a young system in which the rotational and orbital velocities have not yet been equalised by tidal effects.

The reddening of the system has been found using the {\sc uvbybeta} code, which implements the $(b-y)_0$ vs.\ $c_0$ relation by Crawford (\cite{crawford78}). The reddening for the system indices and for the individual stars is ${\Eby}_{\rm \ A+B} = 0.180 \pm 0.007$, ${\Eby}_{\rm \ A} = 0.177 \pm 0.054$ and ${\Eby}_{\rm \ B} = 0.182 \pm 0.064$, where the quoted errors come only from varying the indices within their uncertainties. The main contribution to the reddening uncertainties for the individual stars comes from the uncertainty in their $(b-y)$ colours. We will adopt $\Eby = 0.18 \pm 0.02$ for DW\,Car, where the quoted error includes a contribution from the estimated uncertainty in the calibration.

\subsection{The distance to DW\,Car}

Now we have the absolute dimensions and effective temperatures of the two stars we can calculate the distance to the system. Distances are in general more accurate and precise at infrared wavelengths (Southworth et al.\ \cite{SMSd}), but we are lacking reliable apparent magnitudes in the $JHKL$ passbands. These would be available from 2MASS, but are too faint because they were taken during secondary eclipse. $JHKL$ photometry is available from Tapia et al.\ (\cite{tapia88}) but seems to be too bright compared to the optical colours of DW\,Car. The emission-line nature of DW\,Car suggests that there may be a circumbinary disc. This may affect the infrared colours of the system, resulting in brighter apparent magnitudes. We are therefore unable to use infrared magnitudes of DW\,Car to measure its distance with any confidence.

We have used the $V$ band magnitude and flux ratio of DW\,Car (Table~\ref{table:lc:uvby}) to determine the distance to each star and to the system using several sources of bolometric corrections. We used the {\sc jktabsdim} code\footnote{This can be obtained from \\ {\tt http://www.astro.keele.ac.uk/~jkt/codes.html}}, which calculates distances using several different sources of bolometric corrections (Southworth et al.\ \cite{SMSd}). The available sets of bolometric corrections are calculated using, in general, different radiative parameters for the Sun (Bessell, Castelli \& Plez \cite{bessell98}). It is important to adopt the appropriate values when using bolometric corrections, and improper results can be obtained if this is not done (see discussion in Southworth et al.\ \cite{SMSf}).

Adopting the bolometric correction formula of Flower (\cite{flower96}), we find that the individual distances of the two stars are in acceptable agreement (2877 and 2742 pc), and the overall distance measurement to the system is $2810 \pm 160$\,pc, where the main contributor to the uncertainty comes from the reddening. If we use the theoretically-derived $V$ band bolometric corrections of Bessell et al.\ (\cite{bessell98}) or Girardi et al., (\cite{girardi02}) we find distances to DW\,Car of $2710\pm 140$\,pc and $2720 \pm 140$\,pc, respectively. The agreement between the different distance measurements is acceptable given the uncertainties, but not as good as we were expecting. The components of DW\,Car are hot stars, and transforming their visual absolute magnitudes into bolometric ones requires corrections which are large and relatively uncertain. An improved distance estimate could be found by studying the ultraviolet to optical spectral energy distribution of the system (see Guinan et al.\ \cite{guinan98} and Fitzpatrick \& Massa \cite{fitz}).

Using the $RI$ photometry of DW\,Car from Carraro \& Patat (\cite{carraro01}) and bolometric corrections from Girardi et al.\ (\cite{girardi02}) gives distances of $2890 \pm 130$\,pc and $2710 \pm 110$\,pc, for $R$ and $I$ respectively, where the decreasing importance of the uncertainty in reddening towards longer wavelengths can be seen. However, we will adopt $2810 \pm 160$\,pc as our final distance, found from the $V$ band photometry and the Flower bolometric correction formula, because this is an empirical measurement. The corresponding distance modulus is $12.24 \pm 0.13$\,mag.


\section{Comparison with theoretical models}                                        \label{sec:models}

Graphical comparisons between the predictions of theoretical stellar models and the physical properties of a dEB provide a clear and straightforward interpretation of the agreement or disagreement between theory and observation. The most important plot is of mass versus radius, as these are the two quantities which have been directly measured for the dEB (i.e., without using any calibrations). A second plot can be made because we have measured both the radii and effective temperatures of two stars with known masses. Chosing the ordinate quantity to be mass is again the best approach because the masses of the stars have been directly measured and because model predictions are calculated for specific (initial) masses. The abscissal quantity could be either effective temperature or luminosity (or, equivalently, absolute bolometric magnitude); we prefer the former because this quantity has been measured more directly than luminosity.

In dEBs such as DW\,Car, which are composed of two similar stars, the effective temperatures can be very similar and the errorbars often overlap. In this case, the temperature difference suggested by the individual temperatures (Table~\ref{table:absdim}) is $1400 \pm 1420$\,K, whereas results from the light curve analysis give a much more precise temperature difference of $1380 \pm 420$\,K. This can be represented on a mass--temperature plot by giving full-size errorbars to one star (here star A) and only relative errorbars to the other star (B).

\begin{figure*} \resizebox{\hsize}{!}{\includegraphics{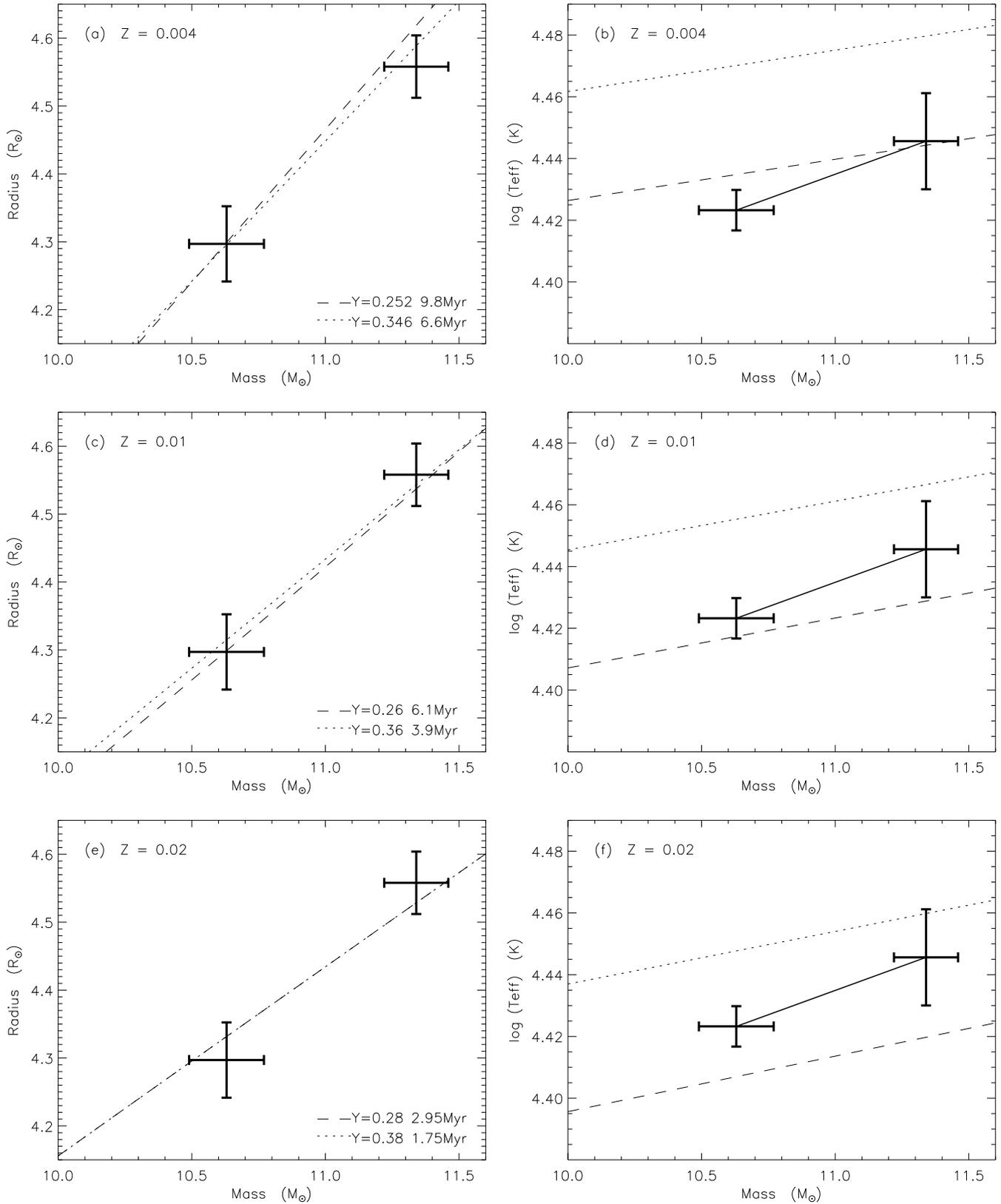}}
\caption{Comparison between the properties of DW\,Car and the predictions of
the Claret stellar evolutionary models. Panels (a) and (b) are a simultaneous
match for metal abundance $Z = 0.004$, panels (c) and (d) likewise for $Z= 0.01$
and panels (e) and (f) for $Z = 0.02$. The model helium abundances and interpolated
ages are given in the legend of each left-hand panel. The errorbars plotted for the
effective temperature of star B are relative to the effective temperature of star A
and {\it do not represent the overall uncertainty}.} \label{fig:model} \end{figure*}

We have chosen to compare the properties of DW\,Car with theoretical predictions from the Claret evolutionary models (Claret \cite{claret95}, \cite{claret97}; Claret \& Gim\'enez \cite{cg95}) which are available for several fractional helium abundances, $Y$, for each fractional metal abundance, $Z$. These models use the opacities of Iglesias, Rogers \& Wilson (\cite{iglesias92}) and incorporate moderate convective core overshooting (with $\alpha_{\rm OV} = 0.20$) and mass loss, and are given for scaled-solar metal abundances.

Isochrones for the best-fitting ages are plotted in Fig.~\ref{fig:model} for metal abundances of $Z = 0.004$, 0.01 and 0.02. A subsolar metal abundance, $Z = 0.01$, provides an excellent match to the masses and radii of DW\,Car, but predictions for the other chemical compositions are also in reasonable agreement here. However, the standard helium abundances in the Claret models (dashed lines) give rise to predicted temperatures slightly too cool for $Z = 0.01$ and $0.02$, although good agreement is found for $Z = 0.004$. A slightly enhanced helium abundance gives a better fit. All models predict shallower mass--\Teff\ slope for DW\,Car, although they agree within the uncertainties.

Other sets of model predictions (Schaller et al.\ \cite{schaller92}; Pols et al.\ \cite{pols98}) were investigated and similar results were found. `Classical' models (without convective core overshooting; Pols et al.\ \cite{pols98}), gave predictions which were negligibly different from their equivalent overshooting models, as expected for such young stars.

Whilst we found the surface helium abundance of DW\,Car to be approximately solar (Section~\ref{sec:spec:He}), the model comparison suggests an overall abundance slightly greater than solar. Note that these two statements are not mutually inconsistent because they refer to different parts of the stars. However, neither abundance measurement is very accurate. We conclude that the best match between stellar models and the properties of DW\,Car is found for a metal abundance of $Z = 0.01$, a helium abundance of $Y = 0.26$ (roughly solar), and an age of 6\,Myr. However, these three quantities are all quite uncertain and have a dependence on theoretical models. The age and chemical composition we infer for DW\,Car are relatively imprecise because its two component stars have similar properties. Much more precise conclusions are possible for dEBs with smaller mass ratios (e.g., Southworth et al.\ \cite{SMSb}).


\section{DW\,Car and Collinder\,228}                                                \label{sec:cluster}

DW\,Car is a photometric member of Cr\,228 and has a sky position close to the cluster centre. However, the systemic velocities of Cr\,228 quoted by Rastorguev et al.\ (\cite{rast99}) and Levato et al.\ (\cite{levato90}), $-12$\kms\ and $-13.5$\kms\ respectively, disagree slightly with our measured systemic velocity for DW\,Car (Table~\ref{table:finalorbit}). However, as noted in Sect.~\ref{sec:finalorbit}, the systemic velocities from the different \'echelle orders are not in full agreement considering their formal errors. Ferrer et al.\ (\cite{ferrer85}) found systemic velocities for the components of DW\,Car in good agreement with the cluster velocity. As our measured distance and reddening are also appropriate for cluster membership, it seems most probable that DW\,Car does belong to Cr\,228. We can therefore derive the important astrophysical properties for Cr\,228 from those for DW\,Car.

The distance we derive, $(m-M)_0 = 12.24 \pm 0.13$\,mag, is in good agreement with most of the measurements in the literature, including that of Th\'e et al.\ (\cite{the80}) for a {\it normal} reddening law, but not including those of Carraro \& Patat (\cite{carraro01}) and Tapia et al.\ (\cite{tapia88}). The disagreement with Tapia et al.\ results from the very large interstellar extinction value they find using infrared photometry. Our distance is the most accurate of the measurements, demonstrating that dEBs are excellent distance indicators. Our reddening value for DW\,Car, equivalent to $\EBV = 0.25 \pm 0.03$\,mag (Crawford \cite{crawford78}), is in reasonable agreement with most literature values.

The age of Cr\,228 is less well known. Determinations on the basis of photometry alone are tricky due to the sparse nature of the cluster, difficulties with measuring reddening, and because $UBV$ photometry is a poor temperature indicator for hot stars (e.g., Massey \& Johnson \cite{massey93}). Massey et al.\ (\cite{massey01}) used classification spectroscopy to measure ages for many stars in Cr\,228 and found an age of about 2\,Myr, in good agreement with our estimate of about 5\,Myr from comparison with the Claret evolutionary models.

The chemical composition of Cr\,228 has not previously been studied. We find a metal abundance of $Z \approx 0.01$ and an approximately solar helium abundance from the properties of DW\,Car. A spectroscopic abundance study of some of the more slowly-rotating members of Cr\,228 would be very useful in specifying the chemical composition of the cluster and of DW\,Car, which would provide an additional constraint when comparing model predictions to the properties of this system.


\section{Summary}                                                             \label{sec:summary}

DW\,Carinae is a young, high-mass detached eclipsing binary system with a short orbital period. It is particularly interesting because it is unevolved, shows Be star emission-line characteristics, and is a member of the young open cluster Collinder 228 in the Carina complex.

The spectra of DW\,Car contain strong hydrogen lines, with some superimposed Be-type emission from circumbinary material, but there are very few other noticable absorption lines. As a result of this, we were only able to measure radial velocities from four or five helium lines. The high rotational velocities of the stars mean that other lines are not detectable, or that small changes in continuum placement can cause a large change in the derived radial velocities. Due to the high rotational velocities, we measured radial velocities using several ways to investigate which are the most reliable in the presence of strong line blending.

Individual radial velocities were obtained by fitting double Gaussians, by cross-correlation, and by the two-dimensional cross-correlation algorithm {\sc todcor} (Zucker \& Mazeh \cite{zucker94}) using the spectrum of a standard star as a template. Spectroscopic orbits were also obtained directly (by least-squares fitting) using spectral disentangling (Simon \& Sturm \cite{simon94}), although this procedure is not straightforward due to a profusion of local maxima in the surface of the quality of the fit in parameter space.

We have found that the radial velocities derived from Gaussian fitting and from cross-correlation require substantial corrections for the effects of line blending. These corrections were derived from simulated composite spectra, constructed using a broadened observed template spectrum, for radial velocity offsets appropriate for each observed spectrum. The corrections were the best-behaved for Gaussian fitting, despite the obvious shortcoming of this technique that helium lines do not have strictly Gaussian shapes. However, the corrections for cross-correlation do not seem to have been completely successful, in that the resulting spectroscopic orbits have systematically lower ampitudes that those derived from the other two methods. The line blending problem is more severe for cross-correlation, because both the object and the template spectra are broadened (using an unbroadened template spectrum causes excessive random errors in the resulting radial velocities). Gaussian fitting, on the other hand, explicitly accounts for the presence of blending lines from the other star, so gives results which are reliable. The cross-correlation results could probably be improved by fitting single-lined CCFs to the double-lined CCFs (see Hill \cite{hill93}). The broadening-function approach advocated by Rucinski (\cite{rucinski02}) is also a useful alternative. The poor performance of {\sc todcor} in dealing with line blending arises because it presents no significant advantage over traditional cross-correlation when the component stars have very similar spectral characteristics (P.\ F.\ L.\ Maxted, private communication).

The orbits measured from spectral disentangling were the most internally consistent, and did not require correcting for blending. Disentangling should be innately more appropriate to this situation because it assumes nothing about the spectral characteristics of the stars, and uses (almost) every observed spectrum in a simultaneous least-squares fit. However, it suffers from the presence of many local maxima in the surface of quality of the fit to the observations. The best approach for measuring spectroscopic orbits in the presence of strong line blending is therefore to use spectral disentangling, checked and corroborated by the results of fitting double Gaussian functions. Conducting such analyses for many lines (helped if the spectra have the wide wavelength coverage achievable using \'echelle spectrographs) is very useful as it allows the overall errors to be both reduced and robustly estimated.

The light curves of DW\,Car were modelled using the Wilson-Devinney code. The ratio of the radii of the stars is intrinsically poorly determined by the light curves due to correlations between various photometric parameters, as is often found for systems with deep but not total eclipses. This degeneracy was broken by constraining the light curve solution using a spectroscopically-derived light ratio. We also fitted directly for the limb darkening coefficients of the stars in order to avoid a dependence on theoretically-derived coefficients. The adoption of different limb darkening laws (linear, square-root and logarithmic) makes a negligible difference to the result, and the fitted coefficients are also in good agreement with the predictions from model atmosphere analyses.

The masses and radii of the components of DW\,Car, calculated from the results of the spectroscopic and photometric analyses, are known to accuracies of about 1\%, so this system joins the short list of well-studied dEBs containing components with masses above 10\Msun. The effective temperatures, and interstellar extinction were derived using photometric calibrations and the individual Str\"omgren indices, which were calculated for the two stars from the combined indices and the light ratios found during the light curve analysis.

The properties of DW\,Car can be used to infer the properties of its parent cluster, Cr\,228. The distance modulus of the dEB, calculated using the empirical bolometric corrections of Flower (\cite{flower96}), is $12.24 \pm 0.13$\,mag, which is the most accurate measurement of Cr\,228's distance yet. A comparison between the properties of DW\,Car and predictions from the theoretical evolutionary models of Claret (\cite{claret95}) was also undertaken. The mass--radius and mass--\Teff\ diagrams are the best choice for such a comparison, because they rely on the properties which are most directly measured and are most easily interpreted. A good match to the properties of the dEB is found for an age of roughly 6\,Myr, a metal abundance of $Z \approx 0.01$, and a solar helium abundance. We have modelled the helium lines of both components of DW\,Car, using non-LTE calculations, to find a roughly solar or slightly subsolar helium abundance which is in acceptable agreement with that inferred from comparison with the theoretical models. The age is in good agreement with literature values for Cr.\,228, and the inferred metal and helium abundances are (to our knowledge) the first measurements for this cluster, illustrating the usefulness of detached eclipsing binaries for the investigation of stellar clusters.

\subsection{DW\,Car, a Be star system}

The emission-line nature of the DW\,Car system is of interest both because of its implications for the properties of the system and in the study of classical Be stars. The Be nature of DW\,Car rests only on strong double-peaked emission in H$\alpha$ and a possible small emission in H$\beta$: there is no noticable emission in the helium or metal lines of the system. The presence of double peaks indicates that the inclination of the circumbinary disc is high, which is consistent with the properties of the orbit of the eclipsing stars and the hypothesis that the orbit and disc should have a common orientation.

The rotational velocities of the components of DW\,Car are towards the lower end of the distribution of Be star rotational velocities (e.g., Chauville et al.\ \cite{chauville01}; Zorec, Fr\'emat \& Cidale \cite{zorec05}). The significance of this is not straightforward to quantify, because we have not considered the effect of gravity darkening when measuring line widths (which tends cause the measured rotational velocities to be underestimates), but is reinforced because we have obtained an {\em equatorial} rotational velocity (which is the upper limit of the possible $V \sin i$ values).

The Be phenomenon is thought to arise from the presence of some kind of pulsation in combination with a rotational velocity close to the critical limit for break-up of the star, causing the ejection of material to form a circumstellar disc (Porter \& Rivinius \cite{porter03}), although magnetic fields may play a part. There are several effects which we would expect to see in DW\,Car. Firstly, about 90\% of Be stars earlier than B5 exhibit small-scale variability (Porter \& Rivinius \cite{porter03}). There are no suggestions of this in our light curves of DW\,Car, but we would not have noticed anything with an amplitude significantly smaller than 5\,$m$mag, which is the photometric precision of the $v$ and $b$ light curve data (Table~\ref{table:lc:lc}; Paper\,I). We would also expect that there would be some light contribution from the circumbinary material, but third light is negligible in the light curves. Both of these effects would have an insignificant effect on the light curve solution, so would not affect the reliability of the radius values. In the absence of intrinsic variability, the presence of a disc (which is confirmed by the H$\alpha$ emission) must be caused by either dynamical effects due to the close-binary nature of DW\,Car or by capture of material from the surrounding environment. The unevolved status of DW\,Car is also consistent with the field population of high-mass Be stars (Zorec et al.\ \cite{zorec05}).


\begin{acknowledgements}

A.\ Kaufer, O.\ Stahl, S.\ Tubbesing, and B.\ Wolf kindly observed nearly all the spectra of DW\,Car during the two periods of Heidelberg/Copenhagen guaranteed time at FEROS in 1998 and 1999. We thank H.\ Hensberge for making his improved version of the MIDAS FEROS package available, and both him and L.\ Freyhammer for valuable advice and help during the reduction of the spectra. We are grateful to E.\ Sturm for providing his original disentangling code, and to him and J.\ D.\ Pritchard for modifying it for use on Linux/Unix computer systems. We would also like to thank P.\ F.\ L.\ Maxted and D.\ J.\ Lennon for useful discussions, P.\ F.\ L.\ Maxted for implementing our version of {\sc todcor}, and the (anonymous) referee for providing a timely and useful report.

JS acknowledges financial support from the Instrument Centre for Danish Astrophysics (IDA) in the form of a postdoctoral grant. The project ``Stellar structure and evolution -- new challenges from ground and space observations'', carried out at Aarhus University and Copenhagen University, is supported by the Danish National Science Research Council. Additional support was received from the (former) Danish Board for Astronomical
Research.

The following internet-based resources were used in research for this paper: the NASA Astrophysics Data System; the SIMBAD database operated at CDS, Strasbourg, France; the VizieR service operated at CDS, Strasbourg, France; and the ar$\chi$iv scientific paper preprint service operated by Cornell University.

This publication makes use of data products from the Two Micron All Sky Survey, which is a joint project of the University of Massachusetts and the Infrared Processing and Analysis Center/California Institute of Technology, funded by the National Aeronautics and Space Administration and the National Science Foundation.

\end{acknowledgements}



\begin{thebibliography}{}

\bibitem[2002]{abt02}         Abt, H.\ A., Levato, H., Grosso, M. 2002, ApJ, 573, 359
\bibitem[1975]{andersen75}    Andersen, J. 1975, A\&A, 44, 355
\bibitem[1991]{andersen91}    Andersen, J. 1991, A\&AR, 3, 91
\bibitem[1983]{andersen83}    Andersen, J., Clausen, J.\ V., Nordstr\"om, B., Reipurth, B. 1983, A\&A, 121, 271
\bibitem[1998]{bessell98}     Bessell, M. S., Castelli, F., \& Plez, B. 1998, A\&A, 333, 231
\bibitem[1985]{butler85}      Butler, K., Giddings, J.\ R. 1985,
                                Newsletter on the Analysis of Astronomical Spectra, No.\ 9
\bibitem[1925]{cannon25}      Cannon, A.\ J. 1925, Ann.\ Astron.\ Obs.\ Harvard Coll., 100, 17
\bibitem[2001]{carraro01}     Carraro, G., Patat, F. 2001, A\&A, 379, 136
\bibitem[2004]{cassisi04}     Cassisi, S., 2004, in IAU Coll.\ 193, Variable Stars in the Local Group, eds.\
                                D.\ W.\ Kurtz and K.\ R.\ Pollard, ASP Conf.\ Ser.\ 310, 489
\bibitem[2005]{cassisi05}     Cassisi, S., 2005, in Resolved Stellar Populations (Cancun, Mexico, 2005), in
                                press (astro-ph/0506161)
\bibitem[2001]{chauville01}   Chauville, J., Zorec, J., Ballereau, D., Morrell, N., Cidale, L., Garcia, A.
                                2001, A\&A, 378, 861
\bibitem[1995]{claret95}      Claret, A. 1995, A\&AS, 109, 441
\bibitem[1997]{claret97}      Claret, A. 1997, A\&AS, 125, 439
\bibitem[1998]{claret98}      Claret, A. 1998, A\&AS, 131, 395
\bibitem[2000]{claret00}      Claret, A. 2000, A\&A, 363, 1081
\bibitem[1995]{cg95}          Claret, A., Gim\'enez, A. 1995, A\&AS, 114, 549
\bibitem[2004]{clausen04}     Clausen, J. V. 2004, New Ast. Rev., 48, 679
\bibitem[1991]{clausen91}     Clausen, J.\ V., Gim\'enez, A. 1991, A\&A, 241, 98
\bibitem[2006]{clausen06}     Clausen, J.\ V., Helt, B.\ E., Gim\'enez, A., et al. 2006, A\&A, submitted
\bibitem[1975]{crawford75}    Crawford, D.\ L. 1975, AJ, 80, 955
\bibitem[1978]{crawford78}    Crawford, D.\ L. 1978, AJ, 83, 48
\bibitem[1977]{davis77}       Davis, J., Shobbrook, R.\ R. 1977, MNRAS, 178, 651
\bibitem[1995]{duflot95}      Duflot, M., Figon, P., Meysonnier, N. 1995, A\&AS, 114, 269
\bibitem[2004]{etzel04}       Etzel, P.B. 2004, SBOP: Spectroscopic Binary Orbit Program
                                (San Diego State University)
\bibitem[1976]{fein76}        Feinstein, A., Marraco, H.\ G., Forte, J.\ C. 1976, A\&AS, 24, 389
\bibitem[1985]{ferrer85}      Ferrer, O.\ E., Niemela, V.\ S., M\'endez, R.\ H., Levato, H., Morrell, N.
                                1985, Rev.\ Mex.\ Astron.\ Astroph., 10, 323
\bibitem[1999]{fitz}          Fitzpatrick, E.\ L., Massa, D. 1999, ApJ, 525, 1011
\bibitem[1996]{flower96}      Flower, P.\ J. 1996, ApJ, 469, 355
\bibitem[1952]{gap52}         Gaposchkin, S. 1952, Ann.\ Astron.\ Obs.\ Harvard Coll., 115, 61
\bibitem[1953]{gap53}         Gaposchkin, S. 1953, Ann.\ Astron.\ Obs.\ Harvard Coll., 113, 67
\bibitem[2002]{girardi02}     Girardi, L., Bertelli, G., Bressan, A., et al. 2002, A\&A, 391, 195
\bibitem[2003]{gies03}        Gies, D.\ R. 2003, in IAU Symp.\ 212, A Massive Star Odyssey: From Main Sequence
                                to Supernova, eds.\ K.\ A.\ van der Hucht, A.\ Herrero and C.\ Esteban, p.\,91
\bibitem[1992]{gray92}        Gray, D.\ F. 1992, The observation and analysis of stellar photospheres,
                                Cambridge University Press, p.\,374
\bibitem[1998]{guinan98}      Guinan, E.\ F., Fitzpatrick, E.\ L., DeWarf, L.\ E., et al. 1998, ApJ, 509, L21
\bibitem[2004]{hebb04}        Hebb, L., Wyse, R.\ F.\ G., Gilmore, G., 2004, AJ, 128, 2881
\bibitem[2000]{hensberge00}   Hensberge, H., Pavlovski, K., Verschueren, W. 2000, A\&A, 358, 553
\bibitem[1924]{hertz24}       Hertzsprung, E. 1924, Bull.\ Astron.\ Inst.\ Netherlands, 2, 165
\bibitem[1973]{hilditch73}    Hilditch, R.\ W. 1973, MNRAS, 164, 101
\bibitem[2005]{hilditch05}    Hilditch, R.\ W. 2005, The Observatory, 125, 72
\bibitem[1993]{hill93}        Hill, G. in ASP Conf.\ Ser.\ 38, New frontiers in binary star research,
                                eds.\ J.\ C.\ Leung and I.-S.\ Nha, 127
\bibitem[2000]{hog00}         H{\o}g, E., Fabricius, C., Makarov, V.\ V. 2000, A\&A, 355, L27
\bibitem[1998]{hynes98}       Hynes, R.\ I., Maxted, P.\ F.\ L. 1998, A\&A, 331, 167
\bibitem[1992]{iglesias92}    Iglesias, C.\ A., Rogers, F.\ J., Wilson, B.\ G., 1992, ApJ, 397, 771
\bibitem[1999]{feros99}       Kaufer, A., Stahl, O., Tubbesing, S., et al. 1999, The ESO Messenger, 95, 8
\bibitem[2000]{feros00}       Kaufer, A., Stahl, O., Tubbesing, S., et al. 2000, in Proc.\ SPIE 4008, Optical and
                                IR Telescope Instrumentation and Detectors, eds.\ M.\ Iye \& A.\ F.\ Morwood, 459
\bibitem[1979]{kurucz79}      Kurucz, R.\ L., 1979, ApJS, 40, 1
\bibitem[1993]{kurucz93}      Kurucz, R.\ L., 1993, in Light Curve Modelling of Eclipsing Binary Stars,
                                ed.,  E.\ F.\ Milone, Springer-Verlag, 93
\bibitem[1981]{levato81}      Levato, H., Malaroda, S. 1981, PASP, 93, 714
\bibitem[1990]{levato90}      Levato, H., Malaroda, S., Garcia, B., Morrell, N., Solivella, G. 1990, ApJS, 72, 323
\bibitem[2004]{lyubimkov04}   Lyubimkov, L.\ S., Rostopchin, S.\ I., Lambert, D. L. 2004, MNRAS, 351, 745
\bibitem[1997]{maceroni97}    Maceroni, C.; Rucinski, S. M. 1997, PASP, 109, 782
\bibitem[2001]{massey01}      Massey, P., DeGioia-Eastwood, K., Waterhouse, E. 2001, AJ, 121, 1050
\bibitem[1993]{massey93}      Massey, P., Johnson, J. 1993, AJ, 105, 980
\bibitem[2001]{mayer01}       Mayer, P., Lorenz, R., Drechsel, H., Abseim, A. 2001, A\&A, 366, 558
\bibitem[1985]{moon85}        Moon, T. T. 1985, Commun.\ Univ.\ London Obs.\ No.\,78
\bibitem[1985]{moondw85}      Moon, T. T. Dworetsky, M.\ M, 1985, MNRAS, 217, 305
\bibitem[1993]{napi93}        Napiwotzki, R., Sch\"onberner, D., Wenske, V. 1993, A\&A, 268, 653
\bibitem[2005]{pav05}         Pavlovski, K., Hensberge, H. 2005, A\&A, 439, 309
\bibitem[2006]{pav06}         Pavlovski, K., Holmgren, D.\ E., Koubsk\'y, P., Southworth, J., Yang, S. 2006,
                                in Close Binaries in the 21st Century, eds., P.\ Niarchos, E.\ F. Guinan,
                                S.\ Rucinski, A. Gim\'enez, in press (preprint: astro-ph/0510678)
\bibitem[1967]{petrie67}      Petrie, R.\ M., Andrews, D.\ H., Scarfe, C.\ D., 1967, Proc.\ IAU Symp.\ 30,
                                Determination of Radial Velocities and their Applications,
                                eds.\ A.\ H.\ Batten and J.\ F.\ Heard, 221
\bibitem[1998]{pols98}        Pols, O. R., Schr\"oder, K.-P., Hurley, J. R., Tout, C. A., \& Eggleton, P. P.,
                                1998 MNRAS, 298, 525
\bibitem[1984]{popper84}      Popper, D. M. 1984, AJ, 89, 132
\bibitem[1991]{popper91}      Popper, D. M., Hill, G. 1991, AJ, 101, 600
\bibitem[2003]{porter03}      Porter, J. M., Rivinius, T. 2003, PASP, 115, 1153
\bibitem[1999]{rast99}        Rastorguev, A.\ S., Glushkova, E.\ V., Dambis, A.\ K., Zabolotskikh, M.\ V.
                                1999, Ast.\ Lett., 25, 595
\bibitem[2002]{rucinski02}    Rucinski, S.\ M. 2002, AJ, 124, 1746
\bibitem[1992]{schaller92}    Schaller, G., Schaerer, D., Meynet, G., Maeder, A. 1992, A\&AS, 96, 269
\bibitem[1974]{simkin74}      Simkin, S.\ M. 1974, A\&A, 31, 129
\bibitem[1994]{simon94}       Simon, K.\ P., Sturm, E. 1994, A\&A, 281, 286
\bibitem[2004a]{SMSa}         Southworth, J., Maxted, P. F. L., Smalley, B. 2004a, MNRAS, 349, 547
\bibitem[2004b]{SMSb}         Southworth, J., Maxted, P. F. L., Smalley, B. 2004b, MNRAS, 351, 1277
\bibitem[2004c]{SMSc}         Southworth, J., Zucker, S., Maxted, P. F. L., Smalley, B. 2004c, MNRAS, 355, 986
\bibitem[2005a]{SMSd}         Southworth, J., Maxted, P. F. L., Smalley, B. 2005a, A\&A, 429, 645
\bibitem[2005b]{SMSe}         Southworth, J., Smalley, B., Maxted, P. F. L., Claret, A., Etzel, P.\ B.
                                2005b, MNRAS, 363, 529
\bibitem[2005c]{SMSf}         Southworth, J., Maxted, P. F. L., Smalley, B. 2005c, in Transits of Venus: New
                                Views of the Solar System and Galaxy, IAU Coll.\ 196, ed.\ D.\ W.\ Kurtz, p.\,361
\bibitem[1963]{stromgren63}   Str\"omgren, B. 1963, Quart.\ Journal of the Royal Astron., Soc., 4, 8
\bibitem[1988]{tapia88}       Tapia, M., Roth, M., Marraco, H., Ruiz, M.\ T., 1988, MNRAS, 232, 661
\bibitem[1980]{the80}         Th\'e, P.\ S., Bakker, R., Antalova, A. 1980, A\&AS, 41, 93
\bibitem[2001]{oglegc17}      Thompson, I.\ B., Ka{\l}u\.zny, J., Pych, W., et al. 2001, AJ, 121, 3089
\bibitem[1979]{tonry79}       Tonry, J., Davis, M. 1979, AJ, 84, 1511
\bibitem[2002]{torres02}      Torres, G., Ribas, I., 2002, ApJ, 567, 1140
\bibitem[1997]{torres97}      Torres, G., Stefanik, R.\ P., Andersen, J., et al. 1997, AJ, 114, 2764
\bibitem[1980]{turner80}      Turner, D.\ G. Moffat, A.\ F.\ J. 1980, MNRAS, 192, 283
\bibitem[1939]{vdhvg39}       van den Hoven van Genderen, E. 1939, Bull.\ Astron.\ Inst.\ Netherlands, 9, 339
\bibitem[1993]{vanhamme93}    Van Hamme, W. 1993, AJ, 106, 2096
\bibitem[2003]{vanhamme03}    Van Hamme, W., Wilson, R. E. 2003, in ASP Conf.\ Ser.\ Vol.\ 298,
                                GAIA Spectroscopy: Science and Technology, ed.\ U.\ Moonari, 323
\bibitem[1979]{wilson79}      Wilson, R. E. 1979, ApJ, 234, 1054
\bibitem[1993]{wilson93}      Wilson, R. E. 1993, in ASP Conf.\ Ser.\ Vol.\ 38, New Frontiers in Binary Star
                                Research, eds.\ K.-C.\ Leung \& I.-S.\ Nha, 91
\bibitem[1971]{wilson71}      Wilson, R. E., Devinney, E. J. 1971, ApJ, 166, 605
\bibitem[1967]{wolfe67}       Wolfe, R.\ H., Horak, H.\ G., Storer, N.\ W., 1967, in Modern Astrophysics: A
                                Memorial to Otto Struve, ed.\ M.\ Hack, 251
\bibitem[1990]{zombeck90}     Zombeck, M.\ V. 1990, Handbook of Astronomy and Astrophysics
                                (Second Edition), Cambridge Univ.\ Press
\bibitem[2005]{zorec05}       Zorec, J., Fr\'emat, Y., Cidale, L. 2005, A\&A, 441, 235
\bibitem[1994]{zucker94}      Zucker, S., Mazeh, T., 1994, ApJ, 420, 806

\end{thebibliography}
\end{document}